%% file: main.tex
\documentclass[%
reprint,
 amsmath,amssymb,
 aps,
prd,
]{revtex4-2}

\usepackage{graphicx}
\usepackage{dcolumn}
\usepackage{bm}
\usepackage{hyperref}
\usepackage[mathlines]{lineno}

\input{preamble}
\input{format}
\input{commands}

\begin{document}

\preprint{APS/123-QED}

\title{Beyond \textit{j}=1: Observational Constraints on Almost-$\Lambda$CDM Cosmologies}

\author{Jess Worsley}
\email{wrsjes002@myuct.ac.za}
\affiliation{%
Department of Mathematics and Applied Mathematics, University of Cape Town, 7 University Ave N, 7700, South Africa.
}%

\author{Saikat Chakraborty}%
\thanks{Corresponding author}
\email{saikat.chakraborty@nwu.ac.za}
\affiliation{%
Institute of Research and Development, Duy Tan University, Da Nang 550000, Vietnam.
}%
\affiliation{Faculty of Natural Sciences, Duy Tan University, Da Nang 550000, Vietnam.}
\affiliation{%
Center for Space Research, North-West University, Potchefstroom 2520, South Africa
}%

\author{Peter Dunsby}
\email{peter.dunsby@uct.ac.za}
\affiliation{
Department of Mathematics and Applied Mathematics, University of Cape Town, 7 University Ave N, 7700, South Africa
}%
\affiliation{
Center for Space Research, North-West University, Potchefstroom 2520, South Africa.
}%
\affiliation{South African Astronomical Observatory, Observatory 7925, Cape Town, South Africa.}

\date{\today}

\begin{abstract}
The cosmographic condition $j(z)=1$ provides the kinematical signature of the spatially flat $\Lambda$CDM model independently of any specific dark-energy or modified-gravity theory. We investigate the extent to which current observations permit departures from this condition by considering three phenomenological ``almost-$\Lambda$CDM'' cosmographic closures in which the cosmic jerk differs slightly from unity through a small deformation parameter $\epsilon$. The models are constrained using Markov Chain Monte Carlo analyses of recent DESI baryon acoustic oscillation measurements together with compressed Planck cosmic microwave background likelihoods and the Union3, Pantheon+, and DESY5 Type Ia supernova compilations.

Unlike conventional analyses that infer the expansion history after imposing an explicit parameterization of the dark-energy equation of state, our cosmographic framework reconstructs the expansion history directly from observations, with the effective dark-energy equation of state emerging as a derived quantity rather than an assumed one. We find that all three cosmographic closures are confined by current observations to a remarkably small neighbourhood of the $\Lambda$CDM cosmographic fixed point, with the inclusion of Planck data driving the preferred evolution toward $j_0\simeq1$ and $w_{\rm DE,0}\simeq-1$. Despite their distinct kinematical constructions, the reconstructed dark-energy evolution consistently exhibits a smooth freezing behaviour that remains close to $w=-1$ without crossing the phantom divide.

Model comparison using the Akaike and Bayesian information criteria shows that the almost-$\Lambda$CDM cosmographic models remain statistically competitive with standard dark-energy parameterizations while relying on fewer assumptions regarding the functional form of $w(z)$. Our analysis demonstrates that current observations constrain the cosmic expansion history more robustly than the detailed evolution of the dark-energy equation of state, which remains sensitive to the reconstruction methodology employed. These results highlight the importance of model-independent cosmographic approaches for interpreting current evidence for dynamical dark energy and provide a natural framework for future studies of cosmological perturbations and structure formation.
\end{abstract}

\maketitle


\section{Introduction}
Both the existence of dark energy and its fundamental nature constitute one of the most profound problems in modern cosmology. Observations of Type Ia supernovae (SNe Ia) at the end of the twentieth century first provided compelling evidence that the Universe is currently undergoing a phase of accelerated expansion \citep{riess_observational_1998, perlmutter_measurements_1999}. Subsequent observations of the cosmic microwave background (CMB), baryon acoustic oscillations (BAO), and large-scale structure have since confirmed this picture with increasing precision. Within the standard cosmological paradigm, this acceleration is attributed to a cosmological constant $\Lambda$, giving rise to the spatially flat $\Lambda$CDM model, which remains the simplest and most observationally successful description of cosmic evolution to date.

Despite its phenomenological success, however, the $\Lambda$CDM (or Concordance) model is confronted by several conceptual and theoretical difficulties. Among the most notable are the cosmological constant fine-tuning problem and the coincidence problem, both of which have motivated extensive investigations into possible alternatives involving dynamical dark energy or modifications to general relativity. In many such scenarios, the accelerated expansion is driven not by a constant vacuum energy density, but by a time-dependent dark energy component with an equation of state that evolves dynamically according to $w(t)=p(t)/\rho(t) \neq -1$.

One of the simplest and most widely studied dynamical dark energy models is quintessence \citep{ratra_cosmological_1988}, in which a slowly rolling scalar field $Q$ evolves under the influence of a potential $V(Q)$. The scalar field is minimally coupled to gravity and interacts with the remaining cosmic components only through the standard gravitational sector. Unlike a cosmological constant, quintessence models generally predict a time-dependent equation of state satisfying $-1<w<-1/3$. Extensions involving tracker fields \citep{steinhardt_cosmological_1999} were later introduced to alleviate the coincidence problem by allowing the scalar field energy density to evolve alongside matter over a large range of cosmic history. These tracker solutions possess attractor-like behavior, rendering the late-time evolution relatively insensitive to initial conditions.

More exotic possibilities include phantom dark energy, in which the scalar field possesses a negative kinetic term, leading to an equation of state satisfying $w<-1$. Such models violate the null energy condition, $\rho + p \geq 0$ and predict a future ``big rip'' singularity in which the expansion rate diverges in finite cosmic time \citep{caldwell_phantom_2003}. More general k-essence scenarios introduce non-canonical kinetic terms, allowing for varying sound speeds and richer dynamical behavior \citep{armendariz-picon_essentials_2001}. In addition to scalar-field dark energy models, modifications to general relativity may also generate late-time cosmic acceleration. Of particular interest are $f(R)$ gravity theories, in which the Ricci scalar $R$ in the Einstein--Hilbert action is replaced by a general function $f(R)$.

Distinguishing between these competing scenarios requires increasingly precise observational constraints on the expansion history of the Universe and the evolution of the dark energy equation of state. Important constraints arise from CMB anisotropies measured by \textit{Planck} \citep{aghanim_planck_2020_cmb}, BAO measurements \citep{eisenstein_detection_2005}, and large-scale structure surveys such as the Dark Energy Survey (DES) \citep{abbott_dark_2024}. Recent observational developments, particularly from DESI BAO measurements combined with CMB and supernova data, have renewed interest in the possibility of mildly evolving dark energy. Although current evidence for deviations from $\Lambda$CDM remains statistically inconclusive, these developments motivate the study of cosmological evolutions that remain close to, but not exactly identical with, standard $\Lambda$CDM.

A particularly useful framework for studying such deviations is cosmography. Cosmographic methods describe the expansion history kinematically through a Taylor expansion of the scale factor or Hubble parameter without assuming a specific underlying gravitational theory or dark energy model \citep{dunsby_theory_2016}. The resulting cosmographic parameters include the Hubble parameter $H$, deceleration parameter $q$, jerk parameter $j$, and higher-order quantities such as snap $s$. Of particular importance is the cosmic jerk parameter, since the spatially flat $\Lambda$CDM model is characterized kinematically by the condition $j(z)=1$. This relation forms the basis of the statefinder diagnostic introduced by \citet{alam_exploring_2003}, which provides a convenient means of distinguishing departures from $\Lambda$CDM evolution. Recent work has further clarified the relation between the kinematic condition $j=1$ and the dynamical condition $w_{\rm DE}=-1$, demonstrating that the correspondence is not necessarily one-to-one in more general cosmological scenarios.

In this work, we investigate whether observationally allowed departures from the cosmographic $\Lambda$CDM condition $j=1$ can produce viable late-time cosmological evolutions that remain close to $\Lambda$CDM. To this end, we consider the three phenomenological ``almost-$\Lambda$CDM'' evolutionary models from \cite{chakraborty_dynamical_2025} in which the jerk parameter deviates slightly from unity through a small deformation parameter $\epsilon$. These models are designed to represent different classes of near-$\Lambda$CDM behavior in the statefinder plane, including evolutions that asymptotically approach $\Lambda$CDM either in the distant past or future. Rather than specifying a particular underlying dark energy Lagrangian, we adopt a purely kinematic approach and constrain these models using combinations of DESI BAO data, compressed \textit{Planck} CMB likelihoods, and Type Ia supernova compilations consisting of Union3, Pantheon+, and DESY5.

This manuscript is organized as follows. In Section \ref{sec:cosmography}, we outline the cosmographic framework and review the role of the jerk parameter in characterizing $\Lambda$CDM evolution. The almost-$\Lambda$CDM models are introduced in Section \ref{sec:almostLCDM}. The observational datasets and likelihood construction are discussed in Section \ref{sec:data}, followed by the MCMC parameter estimation procedure in Section \ref{sec:mcmc}. The results of the parameter estimation are presented in Section \ref{sec:results}, and the respective cosmological evolutions are depicted in Section \ref{sec:evolution}. Section \ref{sec:embedding} explores embedding of the kinematical scenarios under consideration within possible physical dark energy models using $w-w'$ analysis. The observational constraints and statistical comparisons are presented in Section \ref{sec:results}, followed by a summarized discussion of our analysis \ref{sec:discussion}

\section{Cosmography}\label{sec:cosmography}
\label{sec:cosmography}

Cosmographic methods allow for the study of cosmic evolution without specifying a particular model of the Universe \textit{a priori} \citep{dunsby_theory_2016}. Cosmographical quantities arise from the Taylor expansion of the Hubble parameter $H \equiv \frac{\dot{a}}{a}$ around the current epoch \cite{MacDevette2025}:
\begin{eqnarray}
q \equiv -\frac{1}{H^2}\frac{\ddot{a}}{a}, \qquad j \equiv \frac{1}{H^3}\frac{\dddot{a}}{a},
\end{eqnarray}
which characterize the deceleration and jerk parameters, respectively. Using these series coefficients allows for a direct, model-independent test of deviations from $\Lambda$CDM \citep{chakraborty_dynamical_2025}. In fact, after DESI DR2, some recent papers (see, e.g., \cite{Roy:2025cxk,Fazzari:2025lzd,Carloni:2025dqt}) have advocated investigating the possibility of a dynamical dark energy through the lens of cosmography, rather than any prior assumption of a dark energy equation of state parameterization.

One of the main advantages is that any cosmological model in General Relativity (GR) can be described by algebraic conditions among the cosmographic parameters. The most generic $\Lambda$CDM model (without ignoring the global spatial curvature), for example, can be characterized by $s+2(q+j)+qj=0$ \cite{dunajski_cosmic_2008}. The spatially flat $\Lambda$CDM model in particular, which is dynamically characterized by a dark energy equation of state $w_{\mathrm{DE}}(z)=-1$, is kinematically described by the well-known cosmographic condition $j(z)=1$. The equivalence between these forms is the basis for the statefinder diagnostic, which ascribes the $j=1$ condition to the $\Lambda$CDM model. A recent work \cite{chakraborty_dynamical_2025} pointed out that these kinematical and dynamical descriptions may not necessarily have a one-to-one correspondence, and showed that the condition $j(z)=1$ is equivalent to $w_{\mathrm{DE}}(z)=-1$ only if it is assumed that $w_{\mathrm{DE}}(0)=-1\Leftrightarrow\Omega_{m0}=\frac{2}{3}(1+q_0)$. It was also shown in \cite{chakraborty_dynamical_2025} that the equivalence of the conditions $j(z)=1$ and $w_{\rm DE}(z)=-1$, whenever it holds, is not robust with respect to small deviations from the condition $j(z)=1$.

The current article serves as a follow-up on the findings of \cite{chakraborty_dynamical_2025}, by constraining the cosmographic parameters, and accordingly the small deviation parameter that parameterizes the deviation from the exact $\Lambda$CDM kinetics $j(z)=1$, with BAO, CMB, and SNe Ia data. This allows us to determine if an almost-$\Lambda$CDM model corresponds to an almost-$\Lambda$CDM evolution, i.e. $j(z)\approx1$ -- as opposed to an exact $\Lambda$CDM evolution $j(z)=1$ -- and subsequently to an almost-$\Lambda$CDM fluid dark energy model ($w_{\mathrm{DE}}(z)\approx-1$) \citep{luongo_unified_2014}.

\section{Almost $\Lambda$CDM-like evolutionary models}\label{sec:almostLCDM}

In this section, we introduce the three \emph{almost $\Lambda$CDM-like} phenomenological models that we will consider in this paper. An \emph{exact $\Lambda$CDM-like} cosmic evolution is characterized by the cosmographic condition $j=1$ \citep{dunajski_cosmic_2008, Chakraborty:2022evc}; so we will take three phenomenological models whose corresponding cosmographic conditions deviate slightly from $j=1$, parameterized by a small deviation parameter $\epsilon$:
\begin{itemize}
    \item {\bf Model I:} $j=1+3\epsilon(q+1)$,
    \item {\bf Model II:} $j=1+\epsilon$,
    \item {\bf Model III:} $j=1+3\epsilon(q-1/2)$.
\end{itemize}
The first two models have earlier been considered in \cite{chakraborty_dynamical_2025}. They will be complemented in this paper by the third model, which was first introduced in \cite{chakraborty_theory_2026}.

\begin{figure}
    \centering
    \includegraphics[width=\linewidth]{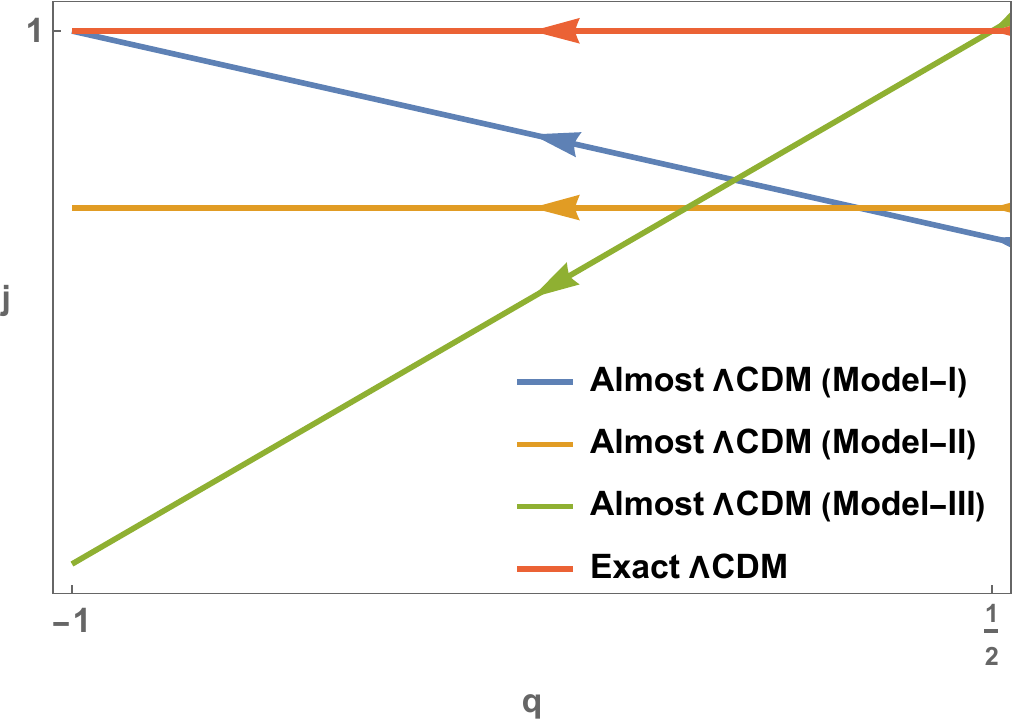}
    \caption{The characteristic evolutionary tracts of the three almost $\Lambda$CDM-like evolutionary models portrayed in the statefinder plane $q-j$. The $\epsilon$-values used to produce these plots are taken from the first columns of Tables \ref{tab:model-I},\ref{tab:model-II},\ref{tab:model-III}, which correspond to DESI DR2 data.}
    \label{fig:statefinder}
\end{figure}

It is helpful to import the notion of the statefinder plane $q-j$ \citep{alam_exploring_2003} to motivate the choice of the three almost $\Lambda$CDM evolutionary models that we have considered here. In the statefinder plane $q-j$, the $\Lambda$CDM model is characterized by the horizontal line $j=1$. The almost $\Lambda$CDM evolutionary model-II is always equidistant from the $\Lambda$CDM model in the statefinder plane, whereas the models I and III coincide with $\Lambda$CDM model in the limit $q\to-1$ and $q\to1/2$, respectively; see Fig.\ref{fig:statefinder}.

Kinematically, the cosmic evolution under model-I becomes asymptotically $\Lambda$CDM in the future, whereas that under model-III becomes asymptotically $\Lambda$CDM in the past, and that under model-II is always close to $\Lambda$CDM but never asymptotes to $\Lambda$CDM. The evolutionary models I and III can be thought of as \emph{kinematic} analogues of freezing and thawing quintessence. We, however, caution that such an analogue does not translate to the actual physical version of freezing or thawing quintessence as we know them (i.e. $w_{\rm DE}\to-1$ in the asymptotic future for freezing quintessence, and $w_{\rm DE}\to-1$ in the asymptotic past for thawing quintessence). As we will see later in \S\ref{sec:embedding}, physically all three models correspond to a freezing-type evolution for the dark energy equation of state. This fact is another example of kinematical and dynamical inequivalence, and one should be cautious about drawing conclusions about the physics of the model from the statefinder diagnostic alone.

\section{Datasets}\label{sec:data}

The DESI collaboration \citep{adame_desi_2025}'s recent combining of their BAO data with CMB data from \textit{Planck} \citep{aghanim_planck_2020_cmb} and ACT \citep{madhavacheril_atacama_2024} delivered indications of an evolving dark energy equation of state \citep{adame_desi_2025}, which was lent credence by the second data release (DR2) in 2025 \citep{collaboration_data_2026, lodha_extended_2025}. Further significance is obtained by combining the BAO and CMB datasets with an assortment of supernovae data: Union3 \citep{rubin_union_2025}, Pantheon+ \citep{scolnic_pantheon_2022, brout_pantheon_2022}, and DESY5 \citep{abbott_dark_2024, sanchez_dark_2024}\footnote{The BAO and SN datasets used in this paper can be found on the \href{https://github.com/CobayaSampler/sn_data/tree/master}{Cobaya GitHub}.}. 

While separate, these datasets have some overlap -- the statistical handling of which is explained in \S\ref{sec:mcmc}. The full combination of BAO, CMB, and SN suggest that $\Lambda$CDM is disfavored at roughly 3-4$\sigma$ compared to a model with evolving dark energy. However, since these results are not independent, it is important to properly interpret the overall level at which $\Lambda$CDM is excluded when considering all the evidence together \citep{cortes_desis_2025}.

\subsection{BAO data}
Baryon acoustic oscillations provide a standard ruler set by the comoving sound horizon at the drag epoch, $r_d$, corresponding to the scale at which baryons decoupled from photons at $z_d \simeq 1060$. Measurements of this scale in the late-time clustering of matter constrain distance ratios to provide robust geometric probes of the expansion history \citep{eisenstein_detection_2005, alam_clustering_2017}.

The Dark Energy Spectroscopic Instrument (DESI) is a Stage IV survey designed to significantly improve cosmological constraints through large-scale structure observations. The full five-year survey covers approximately 14,200 square degrees over the redshift range $0.1 < z < 4.2$, using multiple tracers including luminous red galaxies, emission-line galaxies, quasars, and the Lyman-$\alpha$ forest \citep{adame_desi_2025_lyman}. The wide redshift coverage allows DESI to probe both the late-time acceleration and the transition to matter domination with high precision.

BAO measurements are inherently sensitive to the value of $r_d$, and therefore to early-universe physics. Since our analysis is restricted to late-time evolution, we fix the sound horizon radius to the fiducial value from \textit{Planck}, $r_d\simeq 147.2$ Mpc \citep{aghanim_planck_2020_cosmology}, to avoid introducing an unconstrained degeneracy.

For the DESI BAO likelihood, the model first evaluates the theoretical Hubble expansion rate $H(z)$ on a dense redshift grid in order to compute the cosmological distance measures required by the DESI observations. From the reconstructed expansion history, the Hubble, comoving, transverse comoving, and angular diameter distances are calculated as
\begin{gather*}
    D_H(z) = \frac{c}{H(z)} \,,\\
    \chi(z) = D_H(z)\int_0^z \frac{\dd z'}{\Omega_m(1+z')^3 + \Omega_k(1+z')^2 + \Omega_{\Lambda}} \,,\\
    D_M(z) = \chi(z) \quad\text{for}\quad \Omega_k=0 \,\\
    D_A(z)=\frac{D_M(z)}{1+z} \,,
\end{gather*}
respectively, and the volume-averaged BAO distance used for isotropic measurements is calculated with
\begin{equation*}
    D_V(z)=\left[\frac{(1+z)^2D_A^2(z)cz}{H(z)}\right]^{1/3} \,.
\end{equation*}
The BAO likelihood is a combination of isotropic and anisotropic components $\chi^2_{\text{DESI}} = \chi^2_{\text{iso}} + \chi^2_{\text{aniso}}$, where $\chi^2_{\text{iso}}=\sum_i[A_{\text{th}}(z_i)-A_{\text{obs}}(z_i)]^2/\sigma^2_A(z_i)$ with $A(z)=D_V(z)/r_d$, and $\chi^2_{\text{aniso}}=\Delta D^TC^{-1}_{\text{DESI}}\Delta D$ for residuals in $D_H(z)/r_d$ and $D_M(z)/r_d$ \citep{collaboration_data_2026}. These quantities are interpolated to the DESI redshifts and assembled into a model vector containing the observables measured by DESI, $ \left\{\frac{D_V(z)}{r_s},\frac{D_M(z)}{r_s},\frac{D_H(z)}{r_s}\right\}$,
where $r_s$ is the sound horizon at the drag epoch.

\subsection{SNe Ia data}
Type Ia supernovae provide measurements of the luminosity distance as a function of redshift, making them powerful probes of the expansion history at $z \lesssim 2$. In this work, we consider three major SNe Ia compilations: Union3, Pantheon+, and DESY5.

The Union3 dataset \citep{rubin_union_2025} is an updated compilation of 2087 spectroscopically confirmed SNe Ia at redshifts $0.001<z<2.3$, building on previous Union releases with improved calibration and light-curve fitting. The Pantheon+ sample \citep{scolnic_pantheon_2022, brout_pantheon_2022} is one of the largest and most homogeneous compilations to date, consisting of over 1500 supernovae over the redshift range $0.01 < z < 2.3$, with improved control of systematic uncertainties. The DESY5 sample \citep{abbott_dark_2024, sanchez_dark_2024} is derived from the Dark Energy Survey, spanning redshifts $0.1<z<1.2$ and provides a high-quality, independently calibrated dataset with a well-characterized covariance structure.

The distance moduli of SNe Ia are computed using a modified Tripp relation \citep{tripp_two-parameter_1998, camilleri_dark_2024},
\begin{eqnarray}
\mu_{\text{obs},i} = m_{x,i} + \alpha x_{1,i} - \beta c_i - \gamma G_{\text{host},i} - \mathcal{M} - \Delta\mu_{\text{bias},i} \,,
\end{eqnarray}
where $\mathcal{M}$ absorbs both the absolute magnitude and the Hubble constant. Since $\mathcal{M}$ is degenerate with $H_0$, it cannot be independently constrained by supernova data. The standard approach is therefore to perform analytic marginalization over $\mathcal{M}$ assuming a Gaussian likelihood, rather than sampling the parameter numerically \citep{conley_supernova_2010, betoule_improved_2014, brout_pantheon_2022}.

For a Gaussian likelihood, analytic marginalization leads to a modified inverse covariance matrix \citep{taylor_analytic_2010},
\begin{eqnarray}
C^{-1}_{\text{marginalized}} = C^{-1} - C^{-1} \boldsymbol{1} (\boldsymbol{1}^T C^{-1} \boldsymbol{1})^{-1} \boldsymbol{1}^T C^{-1} \,.
\end{eqnarray}
All supernova datasets were therefore analytically marginalized over the $\mathcal{M}$ parameter \citep{conley_supernova_2010, brout_pantheon_2022}.

While these datasets are often analyzed separately, they are not statistically independent. There is significant overlap in the underlying supernova samples, particularly between Pantheon+ and Union3, and each dataset employs different calibration pipelines and light-curve models. As a result, combining their likelihoods requires careful treatment to avoid double-counting information, as discussed in \citep{cortes_desis_2025}.

The supernova datasets are analyzed with the chi-squared statistic
\begin{eqnarray}
    \chi^2_{\text{SN}} = \Delta D (C_{\text{sys+stat}})^{-1}\Delta D^T \,,
\end{eqnarray}
where $C_{\text{sys+stat}}$ is the (marginalized) combined statistical and systematic covariance matrix, and $\Delta D$ represents the residuals between the observed and theoretical distance moduli for all supernovae. Each is defined as
\begin{eqnarray}
    \Delta D_i = \mu_i - \mu_{\text{th}}(z_i) \,,
\end{eqnarray}
where the observed distance modulus $\mu_i = m_i-M$ is the calculated from the apparent and absolute magnitudes \citep{alam_completed_2021, camilleri_dark_2024, vincenzi_comparing_2025}. For the supernova likelihood, the theoretical Hubble rate $H(z)$ is used to compute the luminosity distance, $D_L(z)=(1+z)c\int_0^z \frac{\dd z'}{H(z')}$. The predicted distance modulus is then 
\begin{eqnarray}
    \mu_{\text{th}}(z,\theta) = 25 + 5\log\left( \frac{d_L(z,\theta)}{1\;\text{Mpc}} \right)\,,
\end{eqnarray}
where $d_L(z,\theta)$ is the luminosity distance and $\theta$ is the set of cosmological parameters \citep{kavya_model-independent_2026}. An additional normalization shift is applied, $\mu \rightarrow \mu - 5\log_{10}\left(\frac{H_0}{70}\right)$, which accounts for the degeneracy between the absolute magnitude calibration and the Hubble constant.

\subsection{CMB data}

The cosmic microwave background provides a snapshot of the Universe at recombination and serves as a cornerstone of modern cosmology. In particular, it tightly constrains early-universe parameters and sets the absolute calibration scale for distance measurements. In this work, we make use of constraints derived from the \textit{Planck} satellite \citep{aghanim_planck_2020_overview}, which provides high-precision measurements of temperature and polarization anisotropies. These measurements constrain parameters such as the baryon density, matter density, and the sound horizon $r_d$, thereby calibrating the BAO distance measurements.

The compressed likelihood is constructed from two distance priors. The shift parameter, $
R = \sqrt{\Omega_{m0}}\frac{H_0\chi_*}{c}$, encodes the scaled comoving distance to recombination, while the acoustic angular scale, $\ell_A = \frac{\pi \chi_*}{r_s(z_*)}$,
measures the angular size of the sound horizon at recombination. These quantities are assembled into the \textit{Planck} model vector and compared with the corresponding observational constraints.

These compressed parameters efficiently capture the information from the CMB that is relevant to late-time cosmological constraints. The \textit{Planck} likelihood is calculated similarly to those above \citep{wang_observational_2007, aghanim_planck_2020_cosmology}: 
\begin{eqnarray}
    \chi^2_{\text{CMB}} = \Delta D (C_{\text{CMB}})^{-1} \Delta D^T \,.
\end{eqnarray}

When combined with low-redshift probes such as BAO and SNe Ia, CMB data play a crucial role in breaking degeneracies (see \S\ref{sec:mcmc}) and enabling precise constraints on the expansion history. However, this combination also introduces dependence on early-universe assumptions, which must be taken into account when interpreting the results.

\section{Parameter estimation with MCMC}\label{sec:mcmc}

In this work, we constrain a set of late-time cosmological parameters $\{H_0, q_0, j_0, \Omega_{m0}\}$, which govern the expansion history at low redshifts. From these values, the quantities $\epsilon, c_1, c_2, w_{\mathrm{DE}}(z)\vert_{z=0}, w'_{\mathrm{DE}}(z)\vert_{z=0} \}$ -- where $\epsilon$, $c_1$, and $c_2$ were introduced in \cite{chakraborty_dynamical_2025} -- are obtained algebraically. The cosmographic parameters $H_0$, $q_0$, and $j_0$ are constrained directly from the observational data. The reconstruction of the dark-energy equation of state additionally requires knowledge of the present matter density parameter $\Omega_{m0}$, which is obtained from the compressed \textit{Planck} distance priors. Therefore the reconstruction of $w_{\text{DE}}$ is not completely model-independent and inherits the assumptions underlying those priors (see Table I of \cite{chen_distance_2019}). The parameter space is constrained using physically motivated, uniform priors; the Hubble constant, for example, is restricted to $50 < H_0 < 72\,\mathrm{km\,s^{-1}\,Mpc^{-1}}$ \citep{Rodrigues:2025tfg}, encompassing values favored by both early- and late-Universe probes while avoiding extreme behaviors that lead to numerical instabilities. These ranges are intentionally broad so that the data, rather than the priors, drive the constraints.

Additional model-dependent conditions are imposed to exclude unphysical regions of parameter space, such as requirements on $j_0$ that enforce a positive energy density and approximate matter domination at high redshift. More generally, parameter combinations that lead to overflow in expressions such as $(1+z)^\alpha$, or produce non-finite or negative values of $H(z)$, $\Omega_m(z)$, or derived distances, are rejected. At the likelihood level, any failure in constructing the model vector results in a log-likelihood of negative infinity, effectively removing unviable regions of parameter space.

The analysis is performed in a reduced parameter space with four free parameters, $\theta = (H_0, q_0, j_0, \Omega_{m0})$. Derived quantities $\epsilon$, $c_1$, $c_2$, $w_{\mathrm{DE}}(z)\vert_{z=0}$, and $w'_{\mathrm{DE}}(z)\vert_{z=0}$ are computed algebraically rather than sampled independently. This reduces the dimensionality of the problem, improving sampling efficiency and convergence while retaining sufficient freedom to capture deviations from $\Lambda$CDM. These derived quantities are calculated using equations \ref{const_redef-I} and \ref{cosmology-I} for model I, \ref{const_redef-II} and \ref{cosmology-II} for model II, and \ref{const_redef-III} and \ref{cosmology-III} for model III. The equation of state parameter and its derivative with respect to redshift at $z=0$ are then found with \citep{chakraborty_dynamical_2025}
\begin{gather}
    w_{\mathrm{DE}}(z)\vert_{z=0} = \frac{1-2q_0}{3(\Omega_{m0}-1)} \,, \\ w'_{\mathrm{DE}}(z)\vert_{z=0} = -  \frac{2j_0 (\Omega_{m0} - 1) + 2q_0(2q_0 - 3\Omega_{m0} + 1) + \Omega_{m0}}{3 (\Omega_{m0} - 1)^2}.
\end{gather}
The likelihood is evaluated in log-space for numerical stability and efficiency, transforming products of probabilities into sums and avoiding underflow \citep{trotta_bayes_2008}. Each dataset contributes a Gaussian likelihood constructed from a model vector and covariance matrix, as outlined in \S\ref{sec:data}. The total likelihood is constructed as the sum of the individual log-likelihoods for the datasets included in a given combination,:
\begin{eqnarray}
\log \mathcal{L}_{\mathrm{tot}} = \log \mathcal{L}_{\mathrm{BAO}} + \log \mathcal{L}_{\mathrm{SN}} + \log \mathcal{L}_{\mathrm{CMB}}.
\end{eqnarray}
This implicitly assumes that the datasets are statistically independent. While this assumption is well justified for combinations involving BAO and CMB data, potential correlations between different supernova compilations should be kept in mind when interpreting combined supernova constraints.

This modular approach allows different dataset combinations to be included consistently \citep{trotta_bayes_2008, lewis_cosmological_2002}. Model vectors provide the link between theory and observations. For a given parameter set, the expansion history $H(z)$ is computed, from which cosmological distances are obtained via numerical integration. These quantities are interpolated to the redshifts of each dataset, and the corresponding observables are assembled into dataset-specific model vectors. This ensures a consistent mapping between parameters and data \citep{hogg_distance_2000, weinberg_observational_2013}.

The MCMC sampling is performed using the \texttt{emcee} affine-invariant ensemble sampler \citep{goodman_ensemble_2010, foreman-mackey_emcee_2013}. An ensemble of 64 walkers is initialized around a fiducial parameter set, followed by a burn-in phase to remove dependence on initial conditions. The chains are then evolved for $\sim10^4$ steps per walker, and convergence is assessed using the integrated autocorrelation time $\tau$. Reliable inference requires chain lengths significantly larger than the integrated autocorrelation time, with effective sample sizes $N_{\mathrm{eff}} \simeq N_{\text{walkers}}N_{\text{steps}}/\tau$. Trace plots are also inspected to verify mixing and stationarity.

We additionally fit the $w_0w_a$CDM model with CPL parameterization \citep{chevallier_accelerating_2001, linder_exploring_2003} using the same assumptions adopted throughout this work, including a fixed sound horizon scale. The resulting constraints should therefore not be interpreted as a reproduction of the DESI collaboration's $w_0w_a$CDM analysis, which marginalizes over the sound horizon through the underlying physical densities, but rather as a benchmark model fitted under the same assumptions as the cosmographic parameterizations.

Finally, model comparison is performed using the Akaike Information Criterion (AIC) and Bayesian Information Criterion (BIC),
\begin{eqnarray}
    \mathrm{AIC} = \chi^2_{\min} + 2k, \qquad \mathrm{BIC} = \chi^2_{\min} + k \ln N,
\end{eqnarray}
where $k$ is the number of free parameters and $N$ is the number of data points \citep{akaike_new_1974, schwarz_estimating_1978, mishra_desi_2026}. These criteria penalize model complexity, allowing different parameterizations to be compared on equal footing, and discouraging overfitting. A model with additional parameters is only preferred if the improvement in fit justifies the increased complexity \citep{liddle_information_2007}. BIC penalizes extra complexity for large datasets especially.

Comparing models involves simply calculating the difference between the information criterion values:
\begin{eqnarray}
    \Delta\text{IC}_i = \text{IC}_i - \text{IC}_{\text{min}} \,,
\end{eqnarray}
where IC is either AIC or BIC. This means that the best model will have a $\Delta\text{IC}$ value of zero. The usual interpretation scale for the weakest model is: $\Delta\text{IC}<2$ implies that the models have equivalent support; $2\le \Delta \text{IC}<6$ implies weak to moderate support for the best model; $6\le \Delta \text{IC}<10$ implies strong support for the best model; and $\Delta \text{IC}\ge10$ indicates very strong support for the best model. We can also quantify the degree to which certain models are favored using the Akaike weights from \cite{wagenmakers_aic_2004}:
\begin{eqnarray}\label{eq:akaike_weights}
    w_i = \frac{\exp\left( -\Delta\text{AIC}_i/2 \right)}{\sum_j \exp\left( -\Delta\text{AIC}_j/2 \right)}\,.
\end{eqnarray}
These weights can be interpreted as the relative support for each model within the set of models being compared, and satisfy $\sum_i w_i=1$.

\section{Results}\label{sec:results}
MCMC analysis was performed with chain lengths of order $10^4$ steps per walker and 64 walkers. Convergence of the MCMC chains was assessed with the integrated autocorrelation time $\tau$ as implemented in \texttt{emcee}. Acceptance fractions were typically $\gtrsim50\%$, and the resulting effective sample sizes all exceeded $1\times 10^4$, making them sufficient for robust parameter estimation. Convergence was further verified with trace plots and running means, which show no residual drift and indicate that the chains have reached a stationary distribution. We therefore conclude that the MCMC chains are adequately converged for all parameters and dataset combinations considered.

\begin{table*}[t]
\tiny
\centering
\begin{tabular}{c|cccccccccc}
Datasets (incl. DESI) & $H_0$                      & $q_0$                      & $j_0$                     & $\Omega_{m0}$             & $\epsilon$                 & $c_1$                     & $c_2$                     & $w_{\mathrm{DE}}(z)\vert_{z=0}$            & $w'_{\mathrm{DE}}(z)\vert_{z=0}$             & $\chi^2_{\text{red}}$ \\ \hline
DESI & $68.292^{+1.706}_{-1.762}$ & $-0.517^{+0.092}_{-0.078}$ & $0.963^{+0.074}_{-0.100}$ & $0.499^{+0.477}_{-0.473}$ & $-0.026^{+0.056}_{-0.055}$ & $4.049^{+1.562}_{-1.241}$ & $3.003^{+3.342}_{-2.852}$ & -- & -- & 1.082 \\
Union3 & $67.352^{+1.400}_{-1.407}$ & $-0.467^{+0.077}_{-0.069}$ & $0.913^{+0.075}_{-0.093}$ & $0.498^{+0.477}_{-0.474}$ & $-0.054^{+0.046}_{-0.045}$ & $3.327^{+1.076}_{-0.879}$ & $2.643^{+2.804}_{-2.513}$ & -- & -- & 1.288 \\
Pantheon+ & $67.189^{+1.124}_{-1.110}$ & $-0.458^{+0.058}_{-0.055}$ & $0.902^{+0.064}_{-0.074}$ & $0.498^{+0.477}_{-0.473}$ & $-0.060^{+0.037}_{-0.037}$ & $3.200^{+0.794}_{-0.672}$ & $2.588^{+2.641}_{-2.456}$ & -- & -- & 1.032 \\
DESY5 & $67.755^{+1.064}_{-1.048}$ & $-0.489^{+0.053}_{-0.050}$ & $0.934^{+0.057}_{-0.066}$ & $0.498^{+0.477}_{-0.473}$ & $-0.043^{+0.037}_{-0.036}$ & $3.616^{+0.819}_{-0.696}$ & $2.793^{+2.835}_{-2.656}$ & -- & -- & 0.941 \\
SN & $67.173^{+0.849}_{-0.831}$ & $-0.456^{+0.041}_{-0.040}$ & $0.898^{+0.053}_{-0.059}$ & $0.504^{+0.472}_{-0.477}$ & $-0.062^{+0.031}_{-0.030}$ & $3.167^{+0.564}_{-0.496}$ & $2.606^{+2.522}_{-2.466}$ & -- & -- & 0.988 \\
Planck & $69.151^{+1.087}_{-1.083}$ & $-0.564^{+0.034}_{-0.032}$ & $1.013^{+0.007}_{-0.008}$ & $0.300^{+0.010}_{-0.010}$ & $0.010^{+0.007}_{-0.007}$ & $4.947^{+0.589}_{-0.534}$ & $2.083^{+0.119}_{-0.109}$ & $-1.013^{+0.021}_{-0.019}$ & $-0.028^{+0.054}_{-0.053}$ & 0.996 \\
Planck + Union3 & $68.635^{+1.008}_{-1.033}$ & $-0.547^{+0.032}_{-0.030}$ & $1.009^{+0.008}_{-0.008}$ & $0.304^{+0.010}_{-0.009}$ & $0.007^{+0.006}_{-0.006}$ & $4.668^{+0.513}_{-0.478}$ & $2.030^{+0.107}_{-0.100}$ & $-1.004^{+0.020}_{-0.019}$ & $-0.002^{+0.052}_{-0.050}$ & 1.405 \\
Planck + Pantheon+ & $68.242^{+0.954}_{-0.960}$ & $-0.534^{+0.031}_{-0.029}$ & $1.006^{+0.007}_{-0.008}$ & $0.308^{+0.009}_{-0.009}$ & $0.004^{+0.006}_{-0.006}$ & $4.464^{+0.460}_{-0.428}$ & $1.990^{+0.096}_{-0.090}$ & $-0.996^{+0.019}_{-0.018}$ & $0.018^{+0.049}_{-0.048}$ & 1.037 \\
Planck + DESY5 & $68.459^{+0.918}_{-0.918}$ & $-0.541^{+0.029}_{-0.027}$ & $1.008^{+0.007}_{-0.008}$ & $0.306^{+0.009}_{-0.009}$ & $0.006^{+0.006}_{-0.005}$ & $4.577^{+0.449}_{-0.416}$ & $2.012^{+0.096}_{-0.088}$ & $-1.000^{+0.018}_{-0.017}$ & $0.007^{+0.047}_{-0.047}$ & 0.944 \\
Planck + SN & $67.741^{+0.796}_{-0.803}$ & $-0.516^{+0.025}_{-0.024}$ & $1.002^{+0.007}_{-0.007}$ & $0.313^{+0.008}_{-0.008}$ & $0.001^{+0.005}_{-0.005}$ & $4.209^{+0.355}_{-0.331}$ & $1.940^{+0.079}_{-0.074}$ & $-0.985^{+0.017}_{-0.016}$ & $0.045^{+0.042}_{-0.042}$ & 0.992
\end{tabular}
\caption{Results and $\chi^2$ values obtained for each parameter and dataset combination using model I (SN = Union3 + Pantheon+ + DESY5). Dark energy equation of state values from BAO/SN have been excluded from the table due to being poorly constrained by these data.}
\label{tab:model-I}
\end{table*}

\begin{table*}[t]
\tiny
\centering
\begin{tabular}{c|cccccccccc}
Datasets (incl. DESI) & $H_0$                      & $q_0$                      & $j_0$                     & $\Omega_{m0}$             & $\epsilon$                 & $c_1$                     & $c_2$                     & $w_{\mathrm{DE}}(z)\vert_{z=0}$            & $w'_{\mathrm{DE}}(z)\vert_{z=0}$             & $\chi^2_{\text{red}}$ \\ \hline
DESI                 & $68.534^{+1.578}_{-1.590}$ & $-0.499^{+0.165}_{-0.153}$ & $0.900^{+0.293}_{-0.269}$ & $0.491^{+0.482}_{-0.469}$ & $-0.100^{+0.293}_{-0.269}$ & $0.326^{+0.106}_{-0.074}$ & $0.491^{+0.482}_{-0.469}$ & -- & -- & 1.091                 \\
Union3               & $67.660^{+1.313}_{-1.319}$ & $-0.409^{+0.143}_{-0.134}$ & $0.757^{+0.240}_{-0.222}$ & $0.500^{+0.475}_{-0.476}$ & $-0.243^{+0.240}_{-0.222}$ & $0.380^{+0.106}_{-0.076}$ & $0.500^{+0.475}_{-0.476}$ & -- & -- & 1.319                 \\
Pantheon+            & $67.378^{+1.099}_{-1.096}$ & $-0.379^{+0.118}_{-0.113}$ & $0.711^{+0.202}_{-0.187}$ & $0.503^{+0.472}_{-0.478}$ & $-0.289^{+0.202}_{-0.187}$ & $0.401^{+0.090}_{-0.069}$ & $0.503^{+0.472}_{-0.478}$ & -- & -- & 1.033                 \\
DESY5                & $67.880^{+1.038}_{-1.036}$ & $-0.433^{+0.111}_{-0.105}$ & $0.794^{+0.197}_{-0.189}$ & $0.507^{+0.468}_{-0.482}$ & $-0.206^{+0.197}_{-0.189}$ & $0.365^{+0.075}_{-0.058}$ & $0.507^{+0.468}_{-0.482}$ & -- & -- & 0.942                 \\
SN                   & $67.268^{+0.836}_{-0.825}$ & $-0.365^{+0.092}_{-0.089}$ & $0.690^{+0.167}_{-0.158}$ & $0.490^{+0.484}_{-0.466}$ & $-0.310^{+0.167}_{-0.158}$ & $0.410^{+0.070}_{-0.056}$ & $0.490^{+0.484}_{-0.466}$ & -- & -- & 0.988                 \\
Planck               & $69.129^{+1.054}_{-1.070}$ & $-0.576^{+0.041}_{-0.038}$ & $1.046^{+0.032}_{-0.031}$ & $0.300^{+0.010}_{-0.009}$ & $0.046^{+0.032}_{-0.031}$  & $0.287^{+0.024}_{-0.022}$ & $0.300^{+0.010}_{-0.009}$ & $-1.025^{+0.028}_{-0.027}$  & $-0.033^{+0.057}_{-0.057}$    & 0.984                 \\
Planck + Union3      & $68.644^{+1.010}_{-1.005}$ & $-0.557^{+0.039}_{-0.037}$ & $1.032^{+0.030}_{-0.029}$ & $0.304^{+0.010}_{-0.009}$ & $0.032^{+0.030}_{-0.029}$  & $0.298^{+0.024}_{-0.022}$ & $0.304^{+0.010}_{-0.009}$ & $-1.013^{+0.027}_{-0.026}$  & $-0.007^{+0.055}_{-0.054}$    & 1.396                 \\
Planck + Pantheon+   & $68.277^{+0.954}_{-0.943}$ & $-0.541^{+0.036}_{-0.036}$ & $1.022^{+0.028}_{-0.027}$ & $0.308^{+0.009}_{-0.009}$ & $0.022^{+0.028}_{-0.027}$  & $0.308^{+0.022}_{-0.021}$ & $0.308^{+0.009}_{-0.009}$ & $-1.003^{+0.026}_{-0.025}$  & $0.013^{+0.051}_{-0.051}$     & 1.037                 \\
Planck + DESY5       & $68.485^{+0.894}_{-0.907}$ & $-0.550^{+0.034}_{-0.033}$ & $1.028^{+0.027}_{-0.026}$ & $0.306^{+0.009}_{-0.008}$ & $0.028^{+0.027}_{-0.026}$  & $0.302^{+0.021}_{-0.019}$ & $0.306^{+0.009}_{-0.008}$ & $-1.008^{+0.024}_{-0.024}$  & $0.002^{+0.048}_{-0.049}$     & 0.944                 \\
Planck + SN          & $67.770^{+0.809}_{-0.799}$ & $-0.520^{+0.030}_{-0.029}$ & $1.008^{+0.024}_{-0.023}$ & $0.312^{+0.008}_{-0.008}$ & $0.008^{+0.024}_{-0.023}$  & $0.320^{+0.019}_{-0.018}$ & $0.312^{+0.008}_{-0.008}$ & $-0.989^{+0.023}_{-0.022}$  & $0.041^{+0.043}_{-0.044}$     & 0.992                
\end{tabular}
\caption{Results and $\chi^2$ values obtained for each parameter and dataset combination using model II (SN = Union3 + Pantheon+ + DESY5). Dark energy equation of state values from BAO/SN have been excluded from the table due to being poorly constrained by these data.}
\label{tab:model-II}
\end{table*}

\begin{table*}[t]
\tiny
\centering
\begin{tabular}{c|cccccccccc}
Datasets (incl. DESI) & $H_0$                      & $q_0$                      & $j_0$                     & $\Omega_{m0}$              & $\epsilon$                 & $c_1$                     & $c_2$                     & $w_{\mathrm{DE}}(z)\vert_{z=0}$            & $w'_{\mathrm{DE}}(z)\vert_{z=0}$             & $\chi^2_{\text{red}}$ \\ \hline
DESI                  & $68.055^{+2.394}_{-1.827}$ & $-0.483^{+0.120}_{-0.172}$ & $0.803^{+0.511}_{-0.275}$ & $0.504^{+0.472}_{-0.479}$ & $0.067^{+0.115}_{-0.158}$  & $0.297^{+0.018}_{-0.017}$ & $0.504^{+0.472}_{-0.479}$ & -- & -- & 1.013                 \\
Union3                & $67.019^{+1.444}_{-1.177}$ & $-0.408^{+0.062}_{-0.099}$ & $0.628^{+0.255}_{-0.120}$ & $0.499^{+0.476}_{-0.473}$ & $0.137^{+0.056}_{-0.098}$  & $0.299^{+0.018}_{-0.017}$ & $0.499^{+0.476}_{-0.473}$ & -- & -- & 1.177                 \\
Pantheon+             & $67.040^{+1.103}_{-1.008}$ & $-0.408^{+0.054}_{-0.071}$ & $0.625^{+0.188}_{-0.115}$ & $0.501^{+0.474}_{-0.476}$ & $0.138^{+0.052}_{-0.074}$  & $0.298^{+0.017}_{-0.017}$ & $0.501^{+0.474}_{-0.476}$ & -- & -- & 1.030                 \\
DESY5                 & $67.582^{+1.092}_{-1.054}$ & $-0.447^{+0.067}_{-0.073}$ & $0.714^{+0.213}_{-0.167}$ & $0.502^{+0.473}_{-0.477}$ & $0.101^{+0.070}_{-0.077}$  & $0.298^{+0.017}_{-0.017}$ & $0.502^{+0.473}_{-0.477}$ & -- & -- & 0.940                 \\
SN                    & $67.091^{+0.836}_{-0.813}$ & $-0.409^{+0.043}_{-0.051}$ & $0.622^{+0.149}_{-0.106}$ & $0.504^{+0.471}_{-0.479}$ & $0.139^{+0.047}_{-0.059}$  & $0.296^{+0.017}_{-0.016}$ & $0.504^{+0.471}_{-0.479}$ & -- & -- & 0.986                 \\
Planck                & $69.872^{+1.715}_{-1.795}$ & $-0.639^{+0.108}_{-0.103}$ & $1.311^{+0.312}_{-0.278}$ & $0.294^{+0.016}_{-0.014}$ & $-0.091^{+0.080}_{-0.076}$ & $0.304^{+0.018}_{-0.016}$ & $0.294^{+0.016}_{-0.014}$ & $-1.075^{+0.081}_{-0.077}$  & $0.052^{+0.020}_{-0.019}$     & 1.399                 \\
Planck + Union3       & $68.212^{+1.407}_{-1.346}$ & $-0.535^{+0.080}_{-0.084}$ & $1.046^{+0.218}_{-0.183}$ & $0.308^{+0.013}_{-0.013}$ & $-0.015^{+0.062}_{-0.064}$ & $0.320^{+0.014}_{-0.014}$ & $0.308^{+0.013}_{-0.013}$ & $-0.998^{+0.062}_{-0.064}$  & $0.052^{+0.019}_{-0.018}$     & 1.548                 \\
Planck + Pantheon+    & $67.592^{+1.096}_{-1.068}$ & $-0.495^{+0.061}_{-0.062}$ & $0.953^{+0.153}_{-0.134}$ & $0.314^{+0.011}_{-0.010}$ & $0.016^{+0.049}_{-0.049}$  & $0.326^{+0.012}_{-0.012}$ & $0.314^{+0.011}_{-0.010}$ & $-0.967^{+0.048}_{-0.049}$  & $0.051^{+0.019}_{-0.018}$     & 1.038                 \\
Planck + DESY5        & $68.108^{+1.057}_{-1.047}$ & $-0.528^{+0.059}_{-0.060}$ & $1.029^{+0.157}_{-0.140}$ & $0.309^{+0.010}_{-0.010}$ & $-0.010^{+0.047}_{-0.048}$ & $0.321^{+0.012}_{-0.011}$ & $0.309^{+0.010}_{-0.010}$ & $-0.992^{+0.047}_{-0.047}$  & $0.051^{+0.019}_{-0.018}$     & 0.946                 \\
Planck + SN           & $67.311^{+0.834}_{-0.828}$ & $-0.476^{+0.044}_{-0.044}$ & $0.910^{+0.110}_{-0.100}$ & $0.317^{+0.009}_{-0.008}$ & $0.031^{+0.037}_{-0.037}$  & $0.329^{+0.010}_{-0.010}$ & $0.317^{+0.009}_{-0.008}$ & $-0.952^{+0.036}_{-0.036}$  & $0.050^{+0.018}_{-0.017}$     & 0.991                
\end{tabular}
\caption{Results and $\chi^2$ values obtained for each parameter and dataset combination using model III (SN = Union3 + Pantheon+ + DESY5). Dark energy equation of state values from BAO/SN have been excluded from the table due to being poorly constrained by these data.}
\label{tab:model-III}
\end{table*}

\begin{figure}[!h]
    \begin{subfigure}[b]{0.92\linewidth}
        \includegraphics[width=\linewidth]{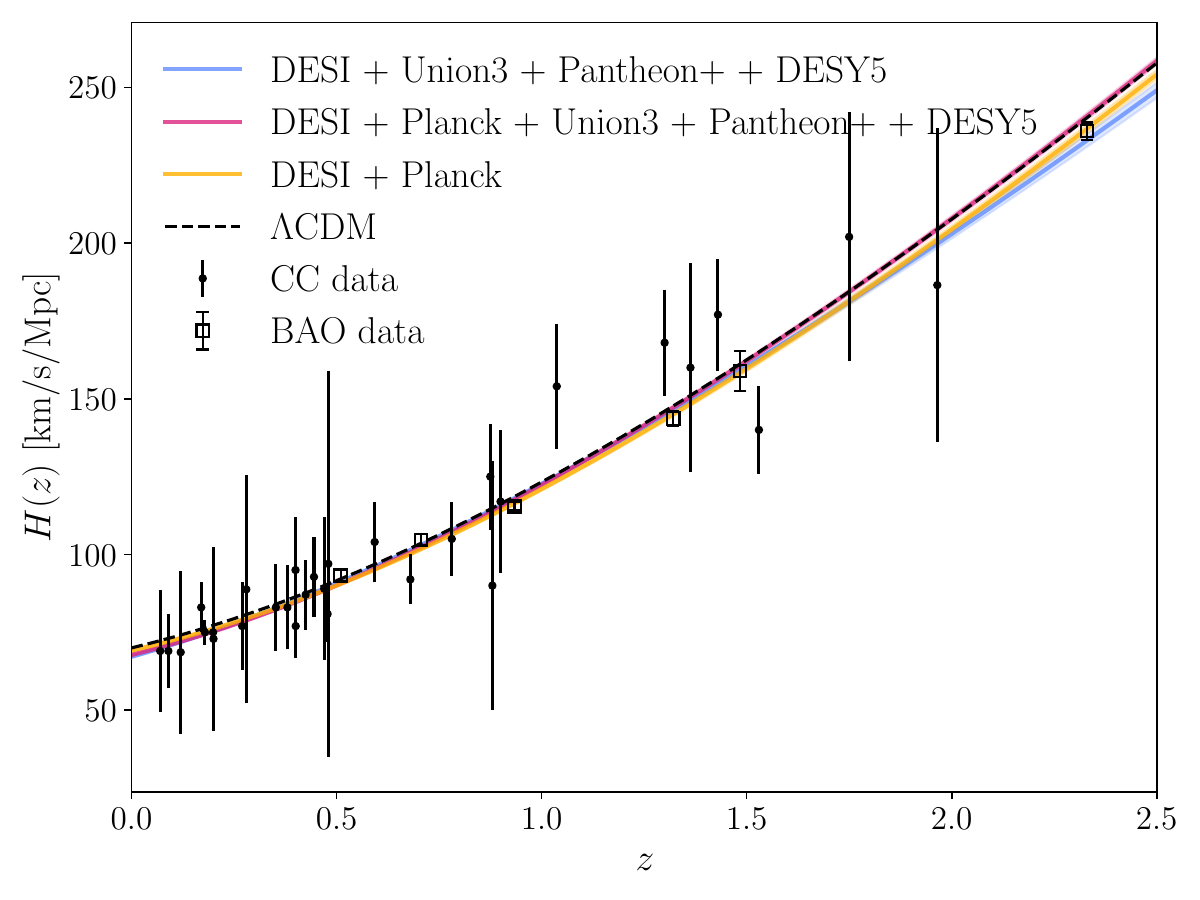}
        \caption{Model I}
    \end{subfigure}
    \begin{subfigure}[b]{0.92\linewidth}
        \includegraphics[width=\linewidth]{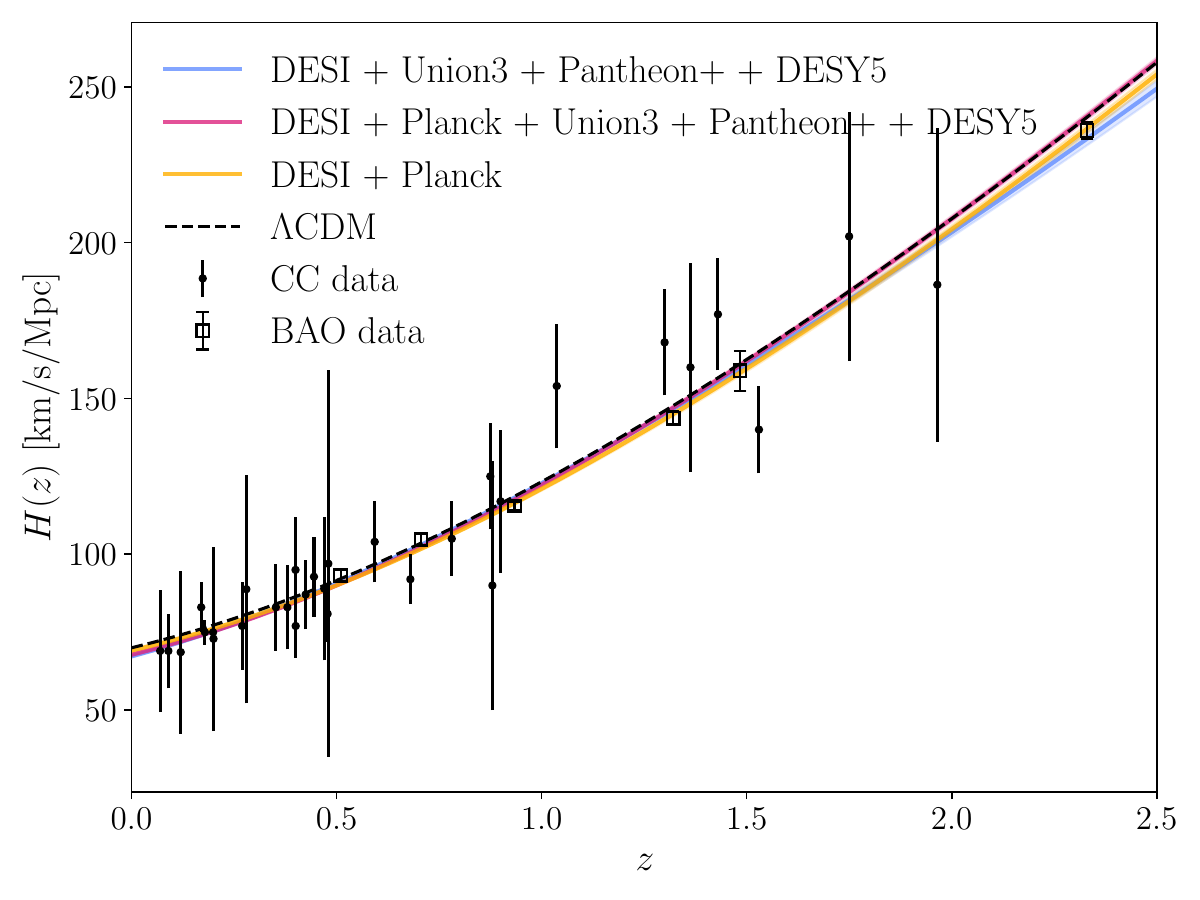}
        \caption{Model II}
    \end{subfigure}
    \begin{subfigure}[b]{0.92\linewidth}
        \includegraphics[width=\linewidth]{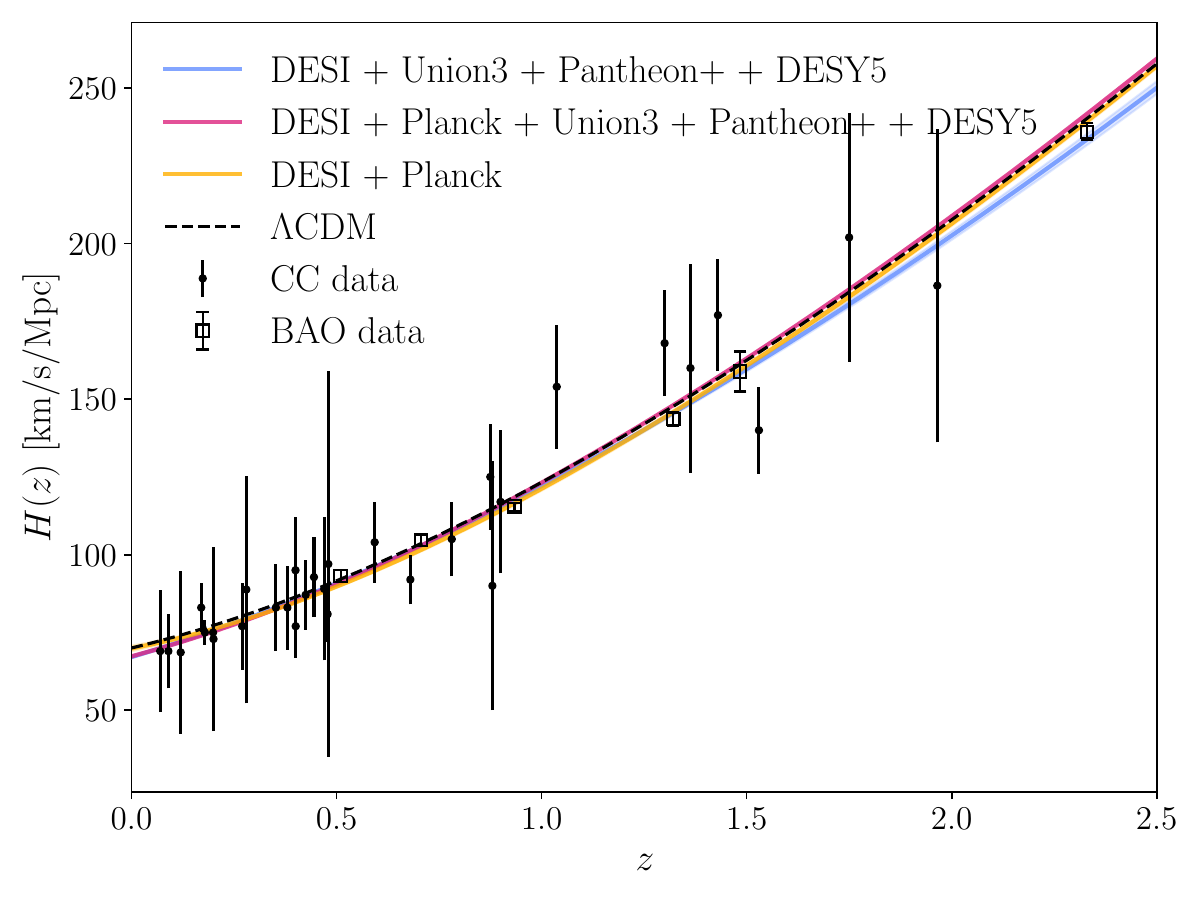}
        \caption{Model III}
    \end{subfigure}
    \caption{Recreated $H(z)$ for each model using DESI + SNIa data (blue), DESI + SNIa + \textit{Planck} data (pink), and DESI + \textit{Planck} data (yellow), alongside $\Lambda$CDM (black dashed line). Cosmic chronometer (black circles) and BAO (black squares) data points have also been included.}
    \label{fig:Hz}
\end{figure}

Estimates of each parameter for models I, II, and III can be found in Tables \ref{tab:model-I}, \ref{tab:model-II}, and \ref{tab:model-III}, respectively. The evolution of $H(z)$ for each model using DESI + SNIa, DESI + SNIa + \textit{Planck}, and DESI + \textit{Planck} is shown alongside $\Lambda$CDM in Fig.\ref{fig:Hz}, and the parameter contours are shown in Fig.\ref{fig:corner}. Reduced chi-squared values for all datasets are $\chi^2_{\text{red}}\approx 1$, indicating satisfactory agreement with the observations. The tightest constraints are obtained when DESI, all supernova samples, and \textit{Planck} data are combined. This combination yields model parameter values of
\begin{equation}
    \epsilon \simeq 0.001^{+0.005}_{-0.005}\, c_1 \simeq 4.209^{+0.355}_{-0.331},\,c_2 \simeq 1.940^{+0.079}_{-0.074}
\end{equation}
for model I,
\begin{equation}
    \epsilon \simeq 0.008^{+0.024}_{-0.023} ,\, c_1 \simeq 0.320^{+0.019}_{-0.018} ,\, c_2 \simeq 0.312^{+0.008}_{-0.008}
\end{equation}
for model II, and
\begin{equation}
    \epsilon \simeq 0.031^{+0.037}_{-0.037} ,\, c_1 \simeq 0.329^{+0.010}_{-0.010} ,\, c_2 \simeq 0.317^{+0.009}_{-0.008}
 \end{equation}
for model III. The evolution of $h(z)$, $q(z)$, and $j(z)$ for these sets of estimated parameters can be seen in Fig.\ref{fig:kinematic}. The AIC results are laid out fully in Table \ref{tab:IC} in Appendix \ref{app:A}, and summarized below. Note that $w$ here refers to the Akaike weight defined in \eqref{eq:akaike_weights}, not the dark energy equation of state.
\begin{align*}
    \Delta\text{AIC}_{\text{I}} &= 2.437, \qquad &w_{\text{I}} &= 0.182, \\
    \Delta\text{AIC}_{\text{II}} &= 2.247, \qquad &w_{\text{II}} &= 0.201, \\
    \Delta\text{AIC}_{\text{III}} &= 0.000, \qquad &w_{\text{III}} &= 0.617,
\end{align*}
indicating that models I and II are moderately disfavored compared to model III. This is also true for the DESI and supernovae data combination, which show substantial support for model III. However, when we exclude supernovae data we find that DESI + \textit{Planck} alone shows more support for models I and II.

\begin{figure}[!h]
    \centering
    \includegraphics[width=0.92\linewidth]{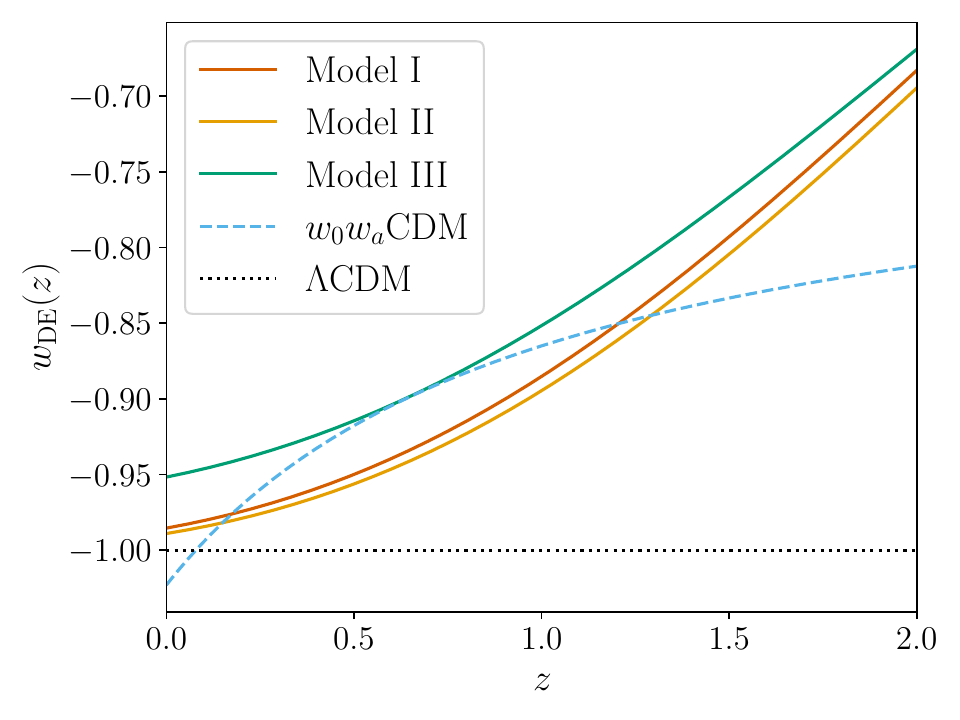}
    \caption{Equation of state parameter $w_{\text{DE}}(z)$ for the DESI + SN + Planck dataset combinations for models I (red), II (yellow), III (green), and $w_0w_a$CDM (blue dashed line), using the values in the final rows of Tables \ref{tab:model-I},\ref{tab:model-II},\ref{tab:model-III}).}
    \label{fig:wz}
\end{figure}

The models were further compared with the CPL parameterization \citep{chevallier_accelerating_2001, linder_exploring_2003} (see Table \ref{tab:model-cpl} in Appendix \ref{app:A}), where $w_0 = w(z=0)$ and $w_a = \dd w / \dd z \vert_{z=0}$. These quantities are directly comparable to the present-day equation-of-state parameter $w_{\text{DE}}(z)\vert_{z=0}$ and its first redshift derivative $w'_{\text{DE}}(z)\vert_{z=0}$ reconstructed in our models. The values obtained by DESI can be found in Table V of \cite{DESI:2025zgx}. When the CPL model is included in the comparison, the Akaike weights favor model III over CPL for the dataset combinations considered here. This conclusion should be interpreted within the context of the specific datasets and assumptions adopted in the present analysis. The equations of state for each model are plotted in Fig. \ref{fig:wz}. Unlike the CPL parameterization, which can favor a phantom crossing, the reconstructed equation-of-state evolutions of models I-III remain consistent with $w_{\text{DE}}(z) \ge -1$ within the observational uncertainties over the redshift range considered.

\section{Cosmic evolution}\label{sec:evolution}

The actual analytical forms of the various cosmological quantities for each of the three models are provided in Appendix \ref{app:B}. Here, we provide the evolutionary plots of these quantities based on the best fit value of the deviation parameter $\epsilon$ for the combination of data sets DESI+Planck+SN (the last columns in Tables \ref{tab:model-I},\ref{tab:model-II},\ref{tab:model-III}).

\begin{figure}[!h]
    \centering    
    \begin{subfigure}{0.88\linewidth}
        \includegraphics[width=\linewidth]{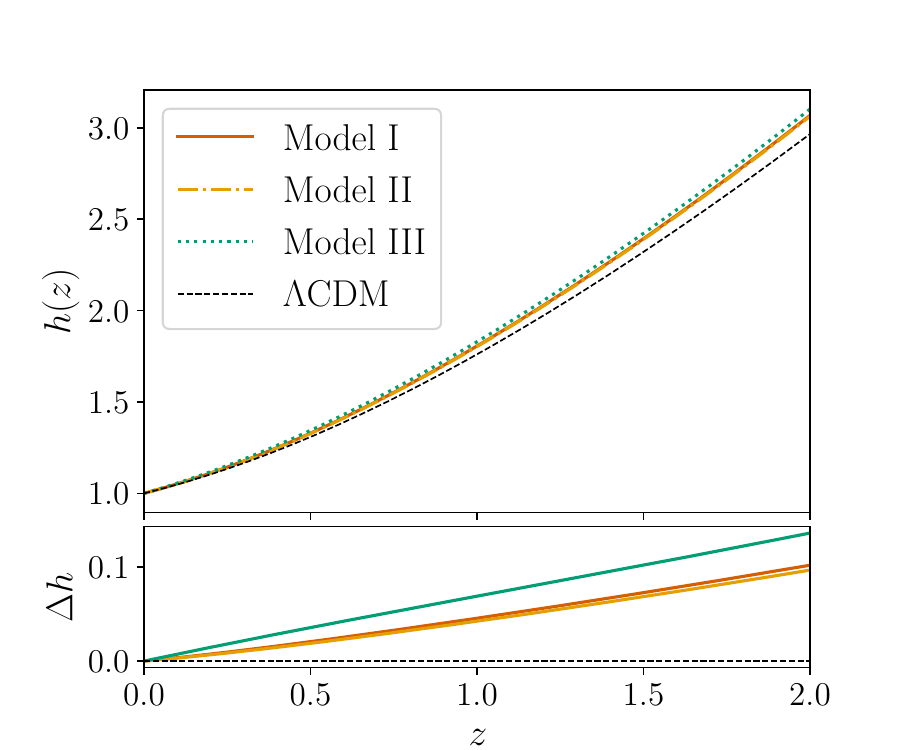}
        \label{fig:hz}
    \end{subfigure}
    \begin{subfigure}{0.88\linewidth}
        \includegraphics[width=\linewidth]{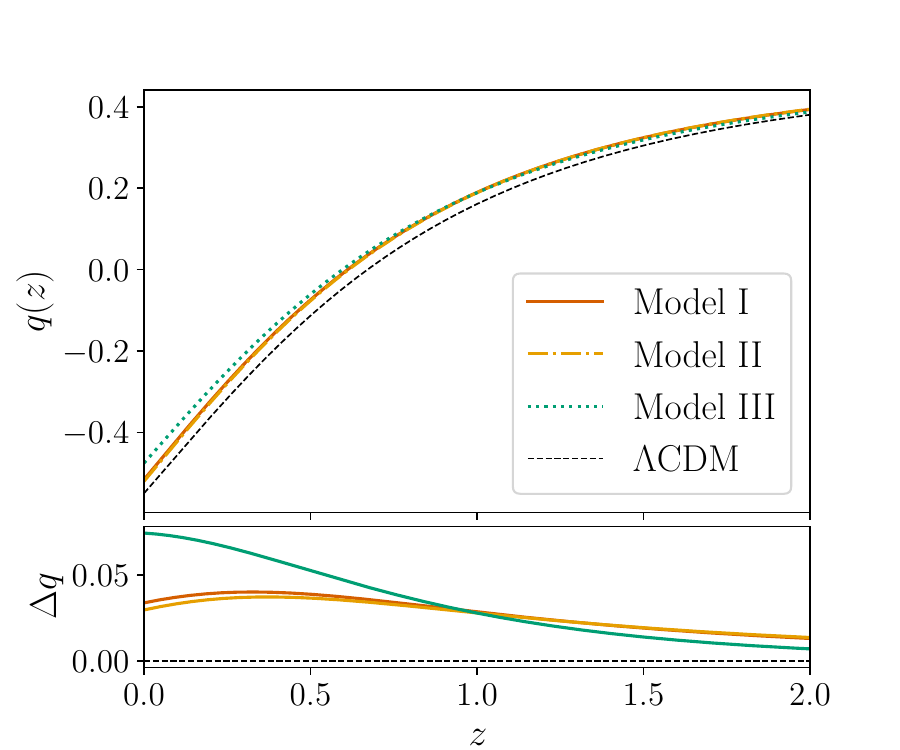}
        \label{fig:qz}
    \end{subfigure}
    \begin{subfigure}{0.88\linewidth}
        \includegraphics[width=\linewidth]{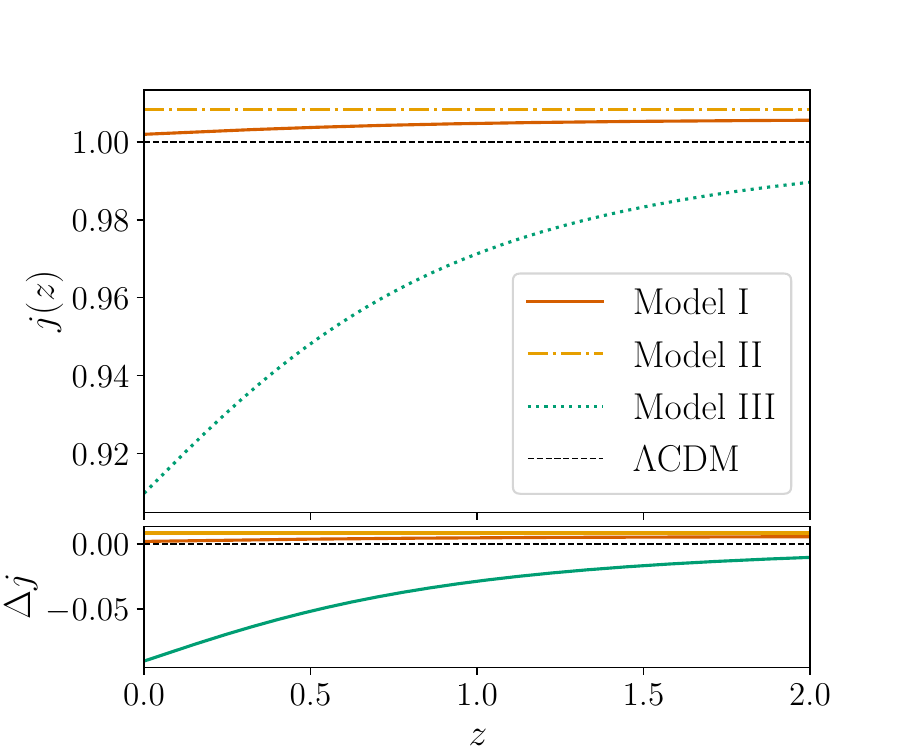}
        \label{fig:jz}
    \end{subfigure}
     \caption{Evolution of kinematic quantities $h$ (top), $q$ (center), and $j$ (bottom) -- and their differences from exact $\Lambda$CDM (black dashed) -- for the almost $\Lambda$CDM evolutionary models I (red solid), II (yellow dash-dotted), III (green dotted line), using the best fit values of $\epsilon$ for DESI+Planck+SN dataset combination (final rows in Tables \ref{tab:model-I},\ref{tab:model-II},\ref{tab:model-III}).}
     \label{fig:kinematic}
\end{figure}

\begin{figure}[!h]
    \begin{subfigure}{\linewidth}
        \includegraphics[width=\linewidth]{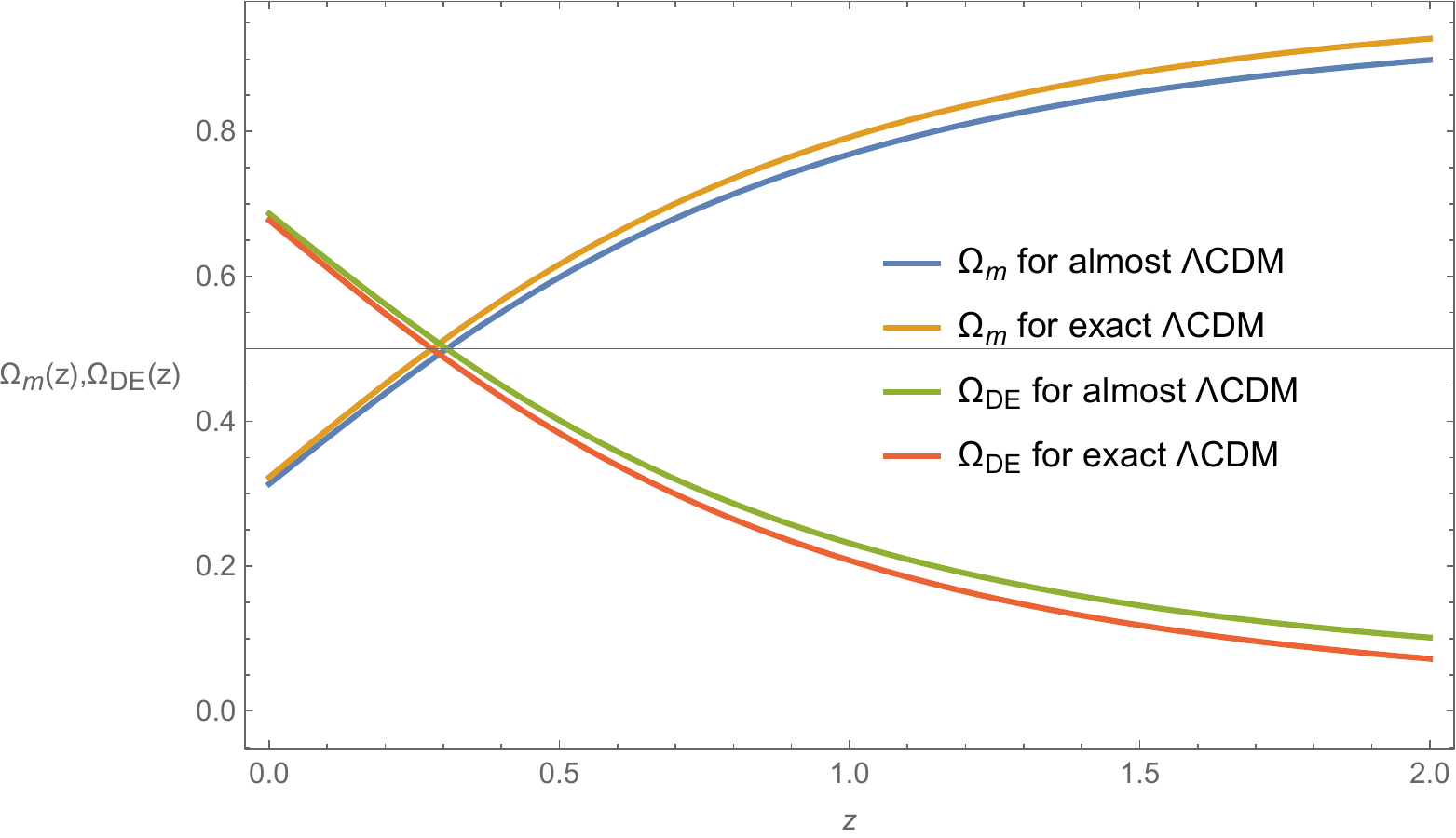}
        \label{fig:Omega1}
    \end{subfigure}
    \hfill
    \begin{subfigure}{\linewidth}
        \includegraphics[width=\linewidth]{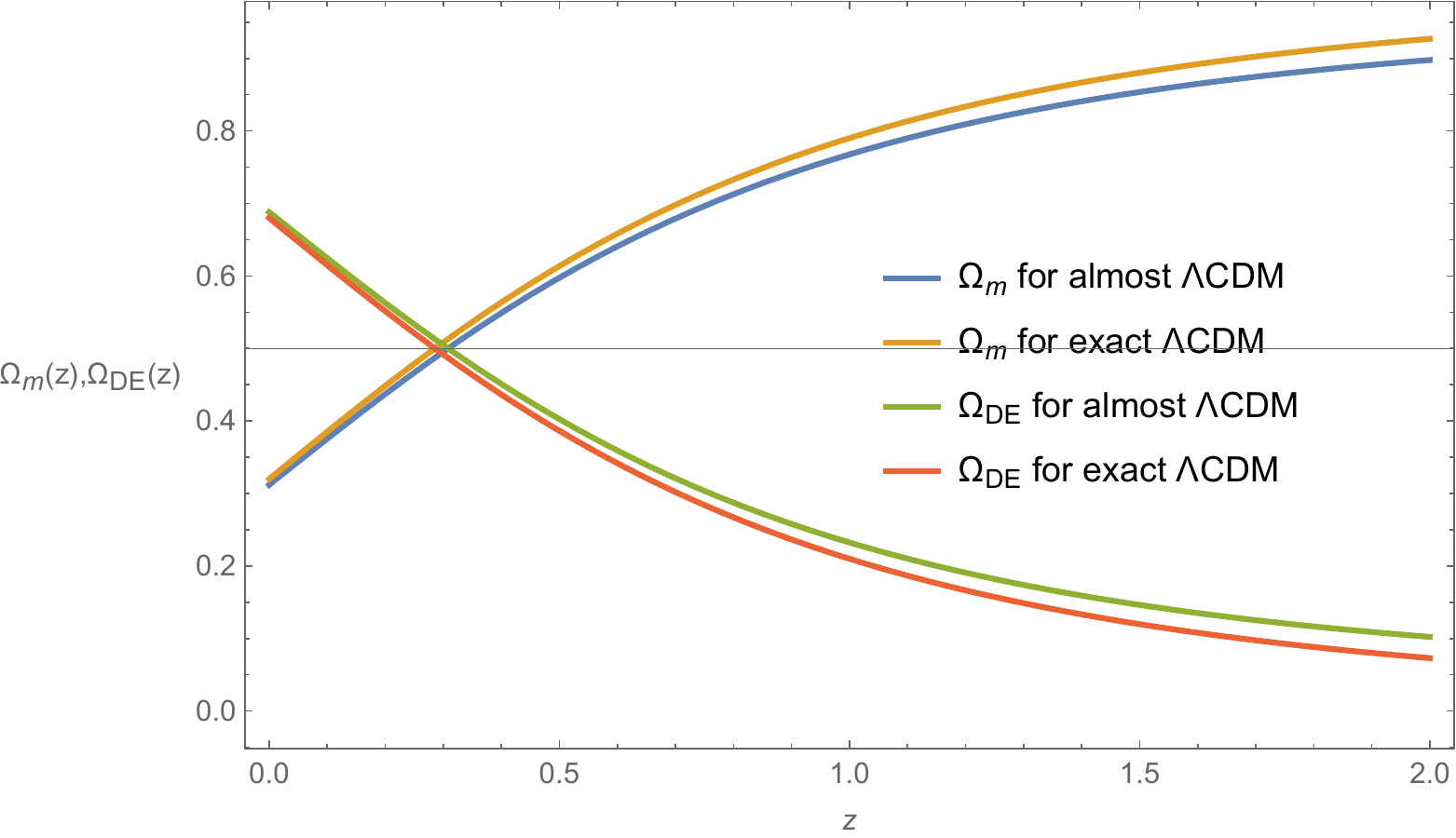}
        \label{fig:Omega2}
    \end{subfigure}
    \hfill
    \begin{subfigure}{\linewidth}
        \includegraphics[width=\linewidth]{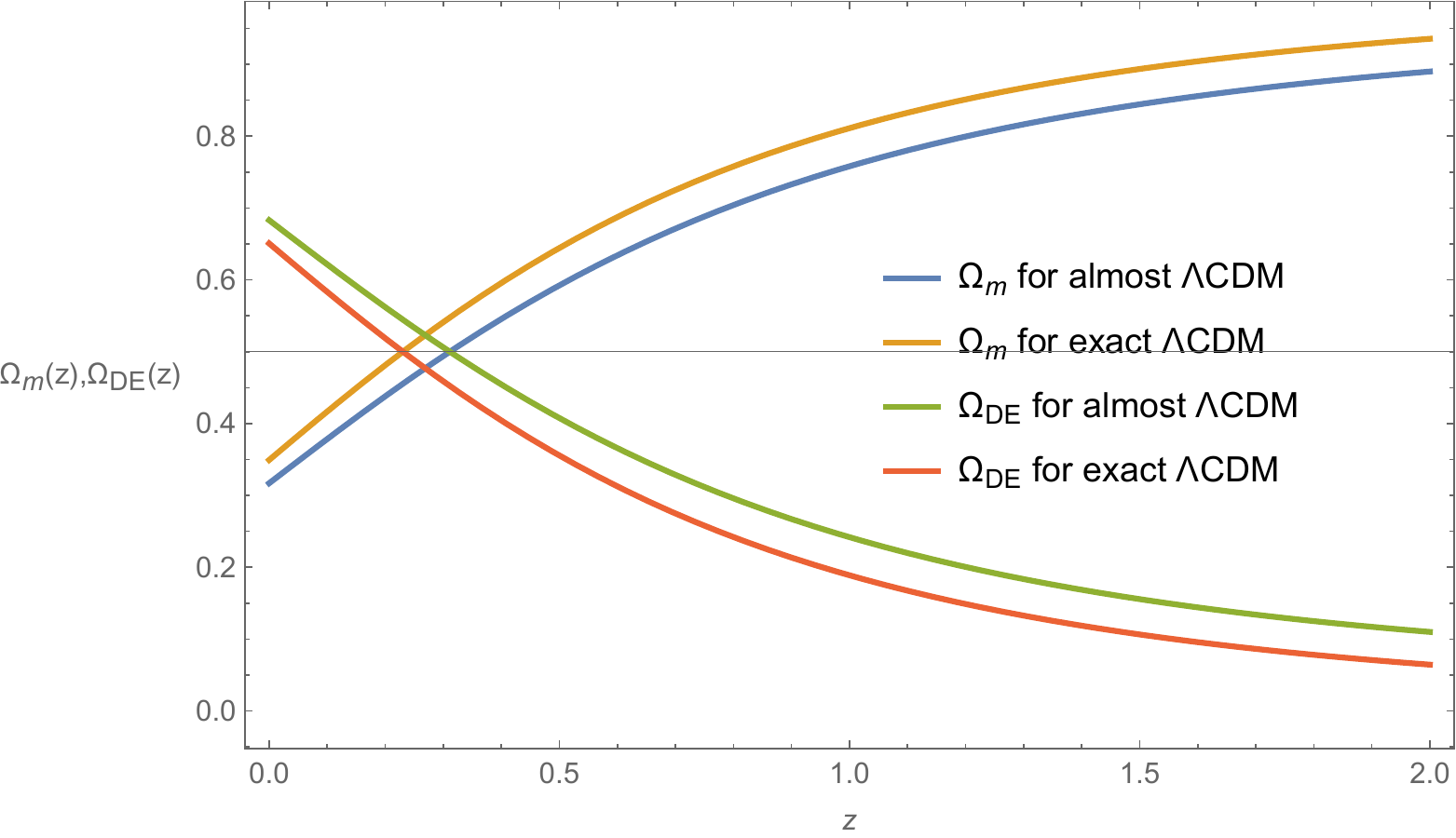}
        \label{fig:Omega3}
    \end{subfigure}
    \caption{Evolution of the density abundance parameters for the almost $\Lambda$CDM evolutionary model-I (left panel), model-II (middle panel), and model-III (right panel). The plots are made for the best fit value of $\epsilon$ for the combination of data sets DESI+Planck+SN (the last columns in Tables \ref{tab:model-I},\ref{tab:model-II},\ref{tab:model-III}).}
    \label{fig:density}
\end{figure}

The plots of the dimensionless Hubble parameter $h(z)$, the deceleration parameter $q(z)$, and the jerk parameter $j(z)$ are shown in Fig.\ref{fig:kinematic}, which, in particular, show that kinematically these models are very close to $\Lambda$CDM. Recognizable differences start appearing when one looks at the evolution of the density abundance parameters (Fig.\ref{fig:density}). Significant differences are apparent when one looks at the evolution of the dark energy equation of state \ref{fig:wz}. The plots in Fig.\ref{fig:wz} also show the freezing nature of the dark energy equation of state ($w_{\rm DE}\to-1$ or to a value close to $-1$ asymptotically in the future). Interestingly, the dark energy equation of state parameter evolution does not show any phantom crossing. We discuss about this observation in \S\ref{sec:discussion}.

\section{Embedding the almost $\Lambda$CDM evolutionary models within possible dynamical dark energy frameworks}\label{sec:embedding}

So far, we have introduced three almost $\Lambda$CDM evolutionary models, constrained their deviation parameters using combinations of data sets, and compared the evolution of various cosmological quantities in these evolutionary models with that of an exact $\Lambda$CDM-like evolution. A natural question is, what kind of physical dark energy models, e.g., quintessence, tachyon, modified gravity, etc., can actually accommodate such almost $\Lambda$CDM-like evolutions. This question is important because cosmographic parameterizations are purely kinematic. Demonstrating that they can be embedded within physically viable dynamical models establishes that the reconstructed evolutions are not merely mathematical parameterizations but correspond to realistic dark-energy scenarios.

In principle, many different classes of dark energy models can admit such cosmic evolutions as solutions, and it might be possible to reconstruct, at least numerically, the particular model belonging to a class (i.e., the particular quintessence potential or the particular $f(R)$ form) that admits a given solution. However, reconstructing a theory from a given solution is a nontrivial task in itself and is beyond the scope of this paper. Below, we present a simple argument showing that, for the best fit value of the deviation parameter $\epsilon$ corresponding to the combination of data sets DESI + \textit{Planck} + SN data, one of the almost $\Lambda$CDM-like evolutions can be physically supported by a freezing quintessence model, whereas all three almost $\Lambda$CDM-like evolutions can be supported by freezing tachyon models.

We will utilize what is known in the literature as $w-w'$ analysis, where $w$ is the dark energy equation of state $w_{\rm DE}$ and $w'=dw/d(\ln a)$. The idea is that, given a dark energy model, it is possible to obtain generic bounds on the value of $w'$ that involve $w$. Consequently, different dark energy models occupy different bands on the $w-w'$ plane \citep{Caldwell:2005tm,Scherrer:2005je,Chiba:2005tj,Chen:2013vba}, although the bands are not always mutually disjoint. In particular, according to \cite{Caldwell:2005tm}, freezing quintessence models occupy a region in the $w-w'$ plane given by
\begin{equation}
  w>-1 \quad \& \quad 3 w (w+1)<w'<0.2 w (w+1)\,,
\end{equation}
whereas, according to \cite{Chen:2013vba}, freezing tachyon models occupy a region in the $w-w'$ plane given by
\begin{equation}
    w>-1 \quad \& \quad -3 (w+1)<w'<0\,.
\end{equation}

\begin{figure}[!h]
    \centering
    \begin{subfigure}{0.45\textwidth}
        \centering
        \includegraphics[width=\linewidth]{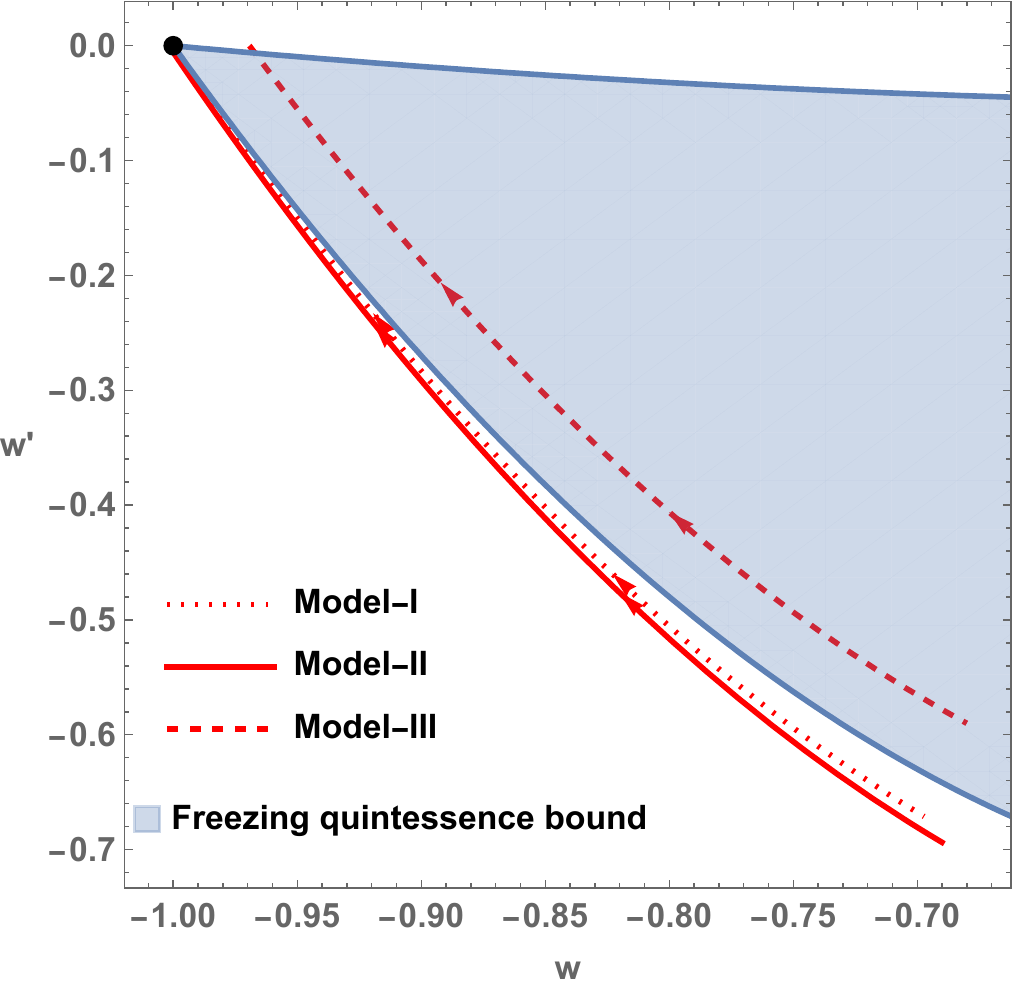}
        \caption{}
        \label{fig:ww'_quintessence}
    \end{subfigure}
    \hfill
    \begin{subfigure}{0.45\textwidth}
        \centering
        \includegraphics[width=\linewidth]{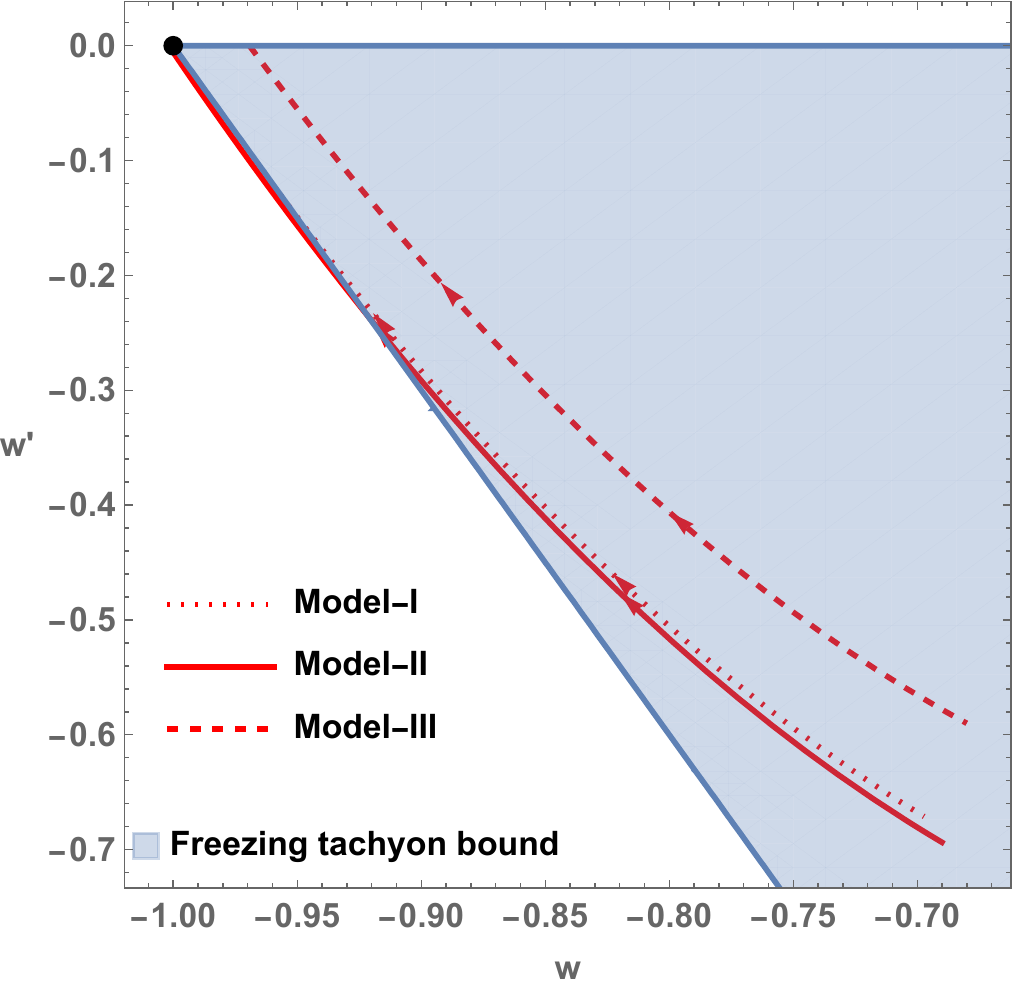}
        \caption{}
        \label{fig:ww'_tachyon}
    \end{subfigure}
    \caption{The $w-w'$ bounds for freezing quintessence models (left panel) and the freezing tachyon model (right panel), as well as the $w-w'$ trajectory for the three almost $\Lambda$CDM evolutionary models.}
    \label{fig:ww'}
\end{figure}

These regions are shown in Fig.\ref{fig:ww'}. On the other hand, with the help of the values of $(\epsilon,c_1,c_2)$, or equivalently $(q_0,j_0,\Omega_{m0})$ listed in Tables \ref{tab:model-I}, \ref{tab:model-II} and \ref{tab:model-III}, the numerical expressions of $w_{\rm DE}(z)$ corresponding to the almost $\Lambda$CDM evolutionary models can be obtained. The latter allows us to obtain the $w-w'$ curve, since $w'=\frac{dw}{d\ln a}=-(1+z)\frac{dw}{dz}$. The respective $w-w'$ curves, for the $\epsilon$-values corresponding to the DESI + \textit{Planck} + SN data (the last columns of Tables \ref{tab:model-I}, \ref{tab:model-II}, \ref{tab:model-III}) are also shown in Fig.\ref{fig:ww'}. 

As can be seen in Fig.\ref{fig:ww'_quintessence}, only the $w-w'$ curve corresponding to the almost $\Lambda$CDM evolutionary model III, namely, the one that kinematically asymptotes to $\Lambda$CDM in the past, falls within the physical bounds of the freezing quintessence model. The other two almost $\Lambda$CDM-like evolutions cannot be supported by a freezing quintessence model. On the other hand, the $w-w'$ curves corresponding to all three almost $\Lambda$CDM evolutions fall within the physical bounds of the freezing tachyon models, which hints at the possibility of realizing all these models within the tachyonic dark energy model framework. The $w-w'$ curves show a freezing-type behavior for all three types of evolutions, although for the model-III, $w_{\rm DE}(z)$ does not actually freezes to a value of $-1$. The $w-w'$ curves for the models are consistent with the $w_{\rm DE}$ evolution shown in Fig.\ref{fig:wz}.  

\section{Discussion}\label{sec:discussion}

In this work we have mapped the region of cosmographic parameter space permitted by current observations around the $\Lambda$CDM fixed point by constraining three distinct almost-$\Lambda$CDM cosmographic closure conditions. Rather than assuming a particular dark-energy model or imposing an explicit parameterization of the dark-energy equation of state, our approach reconstructs the expansion history directly through cosmographic relations, allowing the effective evolution of dark energy to emerge as a consequence of the observational constraints. In this sense, the present analysis provides a genuinely model-independent assessment of how far current observations permit departures from the standard $\Lambda$CDM cosmology.

An important outcome of this work concerns the interpretation of recent claims for evolving dark energy based on DESI observations. The original DESI analyses employed the Chevallier--Polarski--Linder (CPL) parameterization and found statistical evidence for a phantom-to-non-phantom transition around $z\sim0.5$ \cite{DESI:2025zgx}. By contrast, our cosmographic reconstruction makes no {\it a priori} assumption regarding the functional form of $w_{\rm DE}(z)$. Instead, the cosmographic parameters are first constrained directly from the observations, from which the effective dark-energy equation of state is subsequently reconstructed. For the full combination of DESI, DESY5, Union3, Pantheon+, and \textit{Planck} data, this procedure consistently yields a smooth freezing evolution that remains close to $w=-1$ without exhibiting phantom crossing. Our findings therefore indicate that the inferred evolution of the dark-energy equation of state is not uniquely determined by current observations alone, but depends non-trivially on the reconstruction methodology adopted. This conclusion is consistent with other recent cosmographic analyses \cite{Rodrigues:2025tfg,mishra_desi_2026}, which likewise do not require a present-day value $j_0<1$.

Current observations place strong constraints on departures from the cosmographic $\Lambda$CDM condition $j=1$. Although all three almost-$\Lambda$CDM models permit mild deviations from the standard expansion history, the combined observational datasets confine viable cosmological evolution to a remarkably small region of cosmographic parameter space centred on the $\Lambda$CDM fixed point. In particular, the inclusion of \textit{Planck} data drives the preferred solutions towards $j_0\simeq1$ and $w_{\rm DE,0}\simeq-1$, fully consistent with a Universe whose late-time expansion remains extremely close to that of a cosmological constant. While low-redshift probes alone exhibit a modest preference for small departures from $\Lambda$CDM, these become statistically insignificant once the full combination of geometric observations is considered. Correspondingly, model selection using the Akaike and Bayesian Information Criteria provides no significant evidence that the additional phenomenological freedom introduced by the almost-$\Lambda$CDM closures is required by current data.

Despite their distinct kinematical constructions, the three cosmographic models reconstruct similar dark-energy dynamics. In every case the effective equation of state exhibits freezing behaviour, evolving smoothly towards $w=-1$ at late times without crossing the phantom divide. This physically interesting behaviour emerges naturally from the cosmographic reconstruction rather than being imposed through an assumed equation-of-state parameterization. The reconstructed trajectories are compatible with freezing quintessence and freezing tachyon scenarios, demonstrating that relatively small departures from the cosmographic $\Lambda$CDM condition naturally map onto physically viable classes of dynamical dark-energy models.

Several caveats should nevertheless be kept in mind. The present analysis is restricted to the late-time background expansion history and therefore does not include perturbation-level observables such as structure growth, weak gravitational lensing or redshift-space distortions. Since many dynamical dark-energy and modified-gravity models exhibit much stronger signatures at the perturbative level, these observables are expected to provide considerably more stringent tests of almost-$\Lambda$CDM cosmologies than geometric probes alone. Furthermore, the BAO analysis assumes the fiducial \textit{Planck} sound horizon, while compressed CMB likelihoods necessarily inherit assumptions regarding the underlying cosmological model. Residual covariance between some observational datasets may also influence the precise statistical significance of mild departures from $\Lambda$CDM \citep{cortes_desis_2025,vincenzi_comparing_2025,camilleri_dark_2024}. Consequently, the present results should be interpreted primarily as robust constraints on phenomenological departures from the cosmographic condition $j=1$, rather than definitive evidence either for or against dynamical dark energy.

More generally, this work demonstrates that cosmography provides a powerful model-independent framework for investigating the physics of cosmic acceleration. By reconstructing the expansion history directly from observations, rather than assuming a specific parameterization for the dark-energy equation of state, cosmography offers an independent route for interpreting current and future cosmological data. The results presented here show that viable cosmological evolution occupies only a small region of cosmographic parameter space centred on the $\Lambda$CDM fixed point. Future observations from Euclid, the Nancy Grace Roman Space Telescope, LSST, future DESI releases, and complementary probes of cosmological perturbations will determine whether this restricted region reflects the true dynamics of cosmic acceleration or merely the current limits of observational precision. In this sense, the present work represents a first observational mapping of the cosmographic parameter space surrounding the $\Lambda$CDM fixed point and provides a model-independent foundation for future studies of perturbation growth, modified gravity and the physical origin of dark energy.


\section*{Acknowledgments}
JW thanks the University of Cape Town for a Science Faculty PhD scholarship. PKSD thanks the First Rand Bank (SA) for financial support.

\section*{Data and Code Availability}

Data and code used for the MCMC analysis can be found on \href{https://github.com/JessWorsley/LCDM-like-models}{GitHub}.


\newpage
\appendix

\begin{widetext}

\section{Supplementary MCMC results} \label{app:A}

\begin{table}[H]
\centering
\begin{tabular}{c|cccc|cccc|}
                  Datasets (incl. DESI) & \multicolumn{4}{c|}{AIC}                                                                                   & \multicolumn{4}{c|}{BIC}                                                                                   \\ \cline{2-9} 
                   & \multicolumn{1}{c|}{Model I}  & \multicolumn{1}{c|}{Model II} & \multicolumn{1}{c|}{Model III} & CPL Model & \multicolumn{1}{c|}{Model I}  & \multicolumn{1}{c|}{Model II} & \multicolumn{1}{c|}{Model III} & CPL Model \\ \hline
DESI               & \multicolumn{1}{c|}{17.739}   & \multicolumn{1}{c|}{17.817}   & \multicolumn{1}{c|}{17.119}    & \textbf{13.638}    & \multicolumn{1}{c|}{19.999}   & \multicolumn{1}{c|}{20.0768}  & \multicolumn{1}{c|}{19.379}    & \textbf{15.897}    \\
Union3             & \multicolumn{1}{c|}{46.653}   & \multicolumn{1}{c|}{47.582}   & \multicolumn{1}{c|}{43.297}    & \textbf{36.791}    & \multicolumn{1}{c|}{52.759}   & \multicolumn{1}{c|}{53.687}   & \multicolumn{1}{c|}{49.403}    & \textbf{42.897}   \\
Pantheon+          & \multicolumn{1}{c|}{1771.747} & \multicolumn{1}{c|}{1773.353} & \multicolumn{1}{c|}{1767.547}  & \textbf{1763.953}  & \multicolumn{1}{c|}{1793.531} & \multicolumn{1}{c|}{1795.137} & \multicolumn{1}{c|}{1789.331}  & \textbf{1785.737}  \\
DESY5              & \multicolumn{1}{c|}{1651.307} & \multicolumn{1}{c|}{1652.019} & \multicolumn{1}{c|}{1649.123}  & \textbf{1647.064}  & \multicolumn{1}{c|}{1673.177} & \multicolumn{1}{c|}{1673.889} & \multicolumn{1}{c|}{1670.993}  & \textbf{1668.933}  \\
SN                 & \multicolumn{1}{c|}{3431.679} & \multicolumn{1}{c|}{3434.031} & \multicolumn{1}{c|}{3425.968}  & \textbf{3419.626}  & \multicolumn{1}{c|}{3456.288} & \multicolumn{1}{c|}{3458.640} & \multicolumn{1}{c|}{3450.577}  & \textbf{3444.235}  \\
Planck             & \multicolumn{1}{c|}{18.938}   & \multicolumn{1}{c|}{\textbf{18.828}}   & \multicolumn{1}{c|}{23.389}    & 24.493    & \multicolumn{1}{c|}{21.770}   & \multicolumn{1}{c|}{\textbf{21.660}}   & \multicolumn{1}{c|}{26.221}    & 27.326    \\
Planck + Union3    & \multicolumn{1}{c|}{52.949}   & \multicolumn{1}{c|}{\textbf{52.656}}   & \multicolumn{1}{c|}{57.536}    & 65.245    & \multicolumn{1}{c|}{59.283}   & \multicolumn{1}{c|}{\textbf{58.990}}   & \multicolumn{1}{c|}{63.870}    & 71.579    \\
Planck + Pantheon+ & \multicolumn{1}{c|}{1782.758} & \multicolumn{1}{c|}{\textbf{1782.424}} & \multicolumn{1}{c|}{1784.634}  & 1794.885  & \multicolumn{1}{c|}{1804.547} & \multicolumn{1}{c|}{\textbf{1804.213}} & \multicolumn{1}{c|}{1806.422}  & 1816.674  \\
Planck + DESY5     & \multicolumn{1}{c|}{1657.734} & \multicolumn{1}{c|}{\textbf{1657.430}} & \multicolumn{1}{c|}{1661.721}  & 1669.120  & \multicolumn{1}{c|}{1679.608} & \multicolumn{1}{c|}{\textbf{1679.304}} & \multicolumn{1}{c|}{1683.595}  & 1690.994  \\
Planck + SN        & \multicolumn{1}{c|}{3447.940} & \multicolumn{1}{c|}{3447.750} & \multicolumn{1}{c|}{\textbf{3445.503}}  & 3461.335  & \multicolumn{1}{c|}{3472.551} & \multicolumn{1}{c|}{3472.361} & \multicolumn{1}{c|}{\textbf{3470.114}}  & 3485.946 
\end{tabular}
\caption{AIC and BIC scores for each model and dataset combination (SN = Union3 + Pantheon+ + DESY5). The lowest AIC and BIC values for each dataset combination have been written in bold for ease of reading.}
\label{tab:IC}
\end{table}

\begin{table}[H]
\centering
\begin{tabular}{c|ccccc}
Datasets (incl. DESI) & $H_0$                      & $\Omega_{m0}$             & $w_0$                      & $w_a$                      & $\chi^2_{\text{red}}$ \\ \hline
DESI                  & $62.345^{+6.693}_{-4.400}$ & $0.382^{+0.069}_{-0.111}$ & -- & -- & 0.626                 \\
Union3                & $65.039^{+2.058}_{-1.944}$ & $0.347^{+0.032}_{-0.039}$ & -- & -- & 0.960                 \\
Pantheon+             & $66.789^{+1.135}_{-1.129}$ & $0.323^{+0.023}_{-0.034}$ & -- & -- & 1.027                 \\
DESY5                 & $67.261^{+1.195}_{-1.154}$ & $0.318^{+0.025}_{-0.055}$ & -- & -- & 0.951                 \\
SN                    & $66.903^{+0.837}_{-0.831}$ & $0.326^{+0.021}_{-0.026}$ & -- & -- & 0.984                 \\
Planck                & $70.532^{+1.324}_{-1.775}$ & $0.296^{+0.016}_{-0.012}$ & $-1.165^{+0.077}_{-0.065}$ & $0.288^{+0.015}_{-0.011}$  & 1.499                 \\
Planck + Union3       & $68.646^{+1.012}_{-1.009}$ & $0.304^{+0.010}_{-0.009}$ & $-1.087^{+0.070}_{-0.061}$  & $0.304^{+0.015}_{-0.012}$ & 1.789                 \\
Planck + Pantheon+    & $68.275^{+0.954}_{-0.944}$ & $0.308^{+0.009}_{-0.009}$ & $-1.044^{+0.047}_{-0.047}$  & $0.312^{+0.011}_{-0.010}$ & 1.044                 \\
Planck + DESY5        & $68.362^{+1.065}_{-1.015}$ & $0.315^{+0.011}_{-0.013}$ & $-1.070^{+0.046}_{-0.046}$ & $0.307^{+0.010}_{-0.010}$  & 0.950                 \\
Planck + SN           & $67.417^{+0.836}_{-1.882}$ & $0.324^{+0.010}_{-0.024}$ & $-1.023^{+0.385}_{-0.037}$ & $0.316^{+0.032}_{-0.008}$  & 0.995                
\end{tabular}
\caption{Results and $\chi^2$ values obtained for each parameter and dataset combination using the CPL parameterization (SN = Union3 + Pantheon+ + DESY5). Dark energy equation of state values from BAO/SN have been excluded from the table due to being poorly constrained by these data.}
\label{tab:model-cpl}
\end{table}

\begin{figure}[!h]
    \centering
    \begin{subfigure}[b]{0.4\linewidth}
        \includegraphics[width=\linewidth]{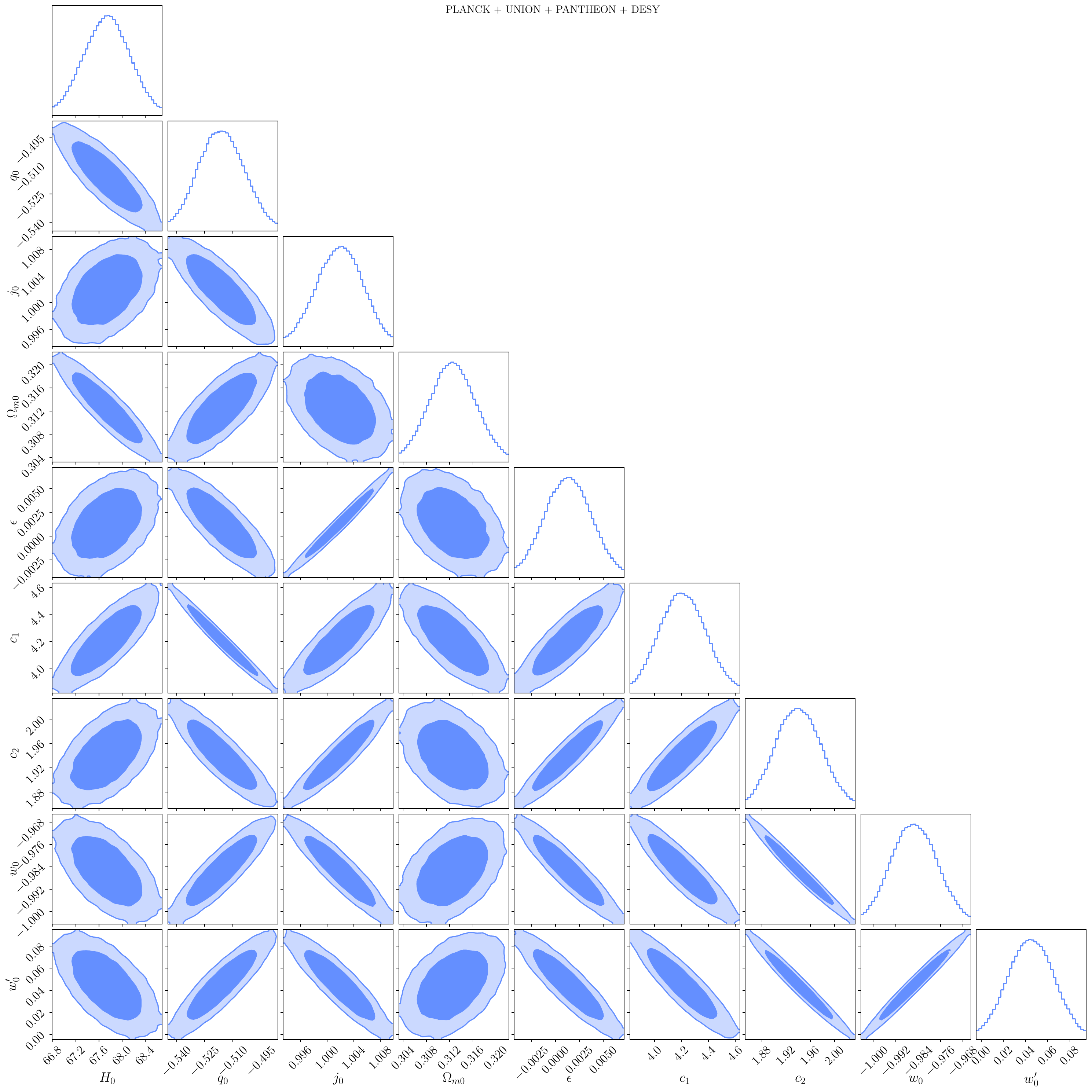}
        \caption{Model I}
    \end{subfigure}
    \hfill
    \begin{subfigure}[b]{0.4\linewidth}
        \includegraphics[width=\linewidth]{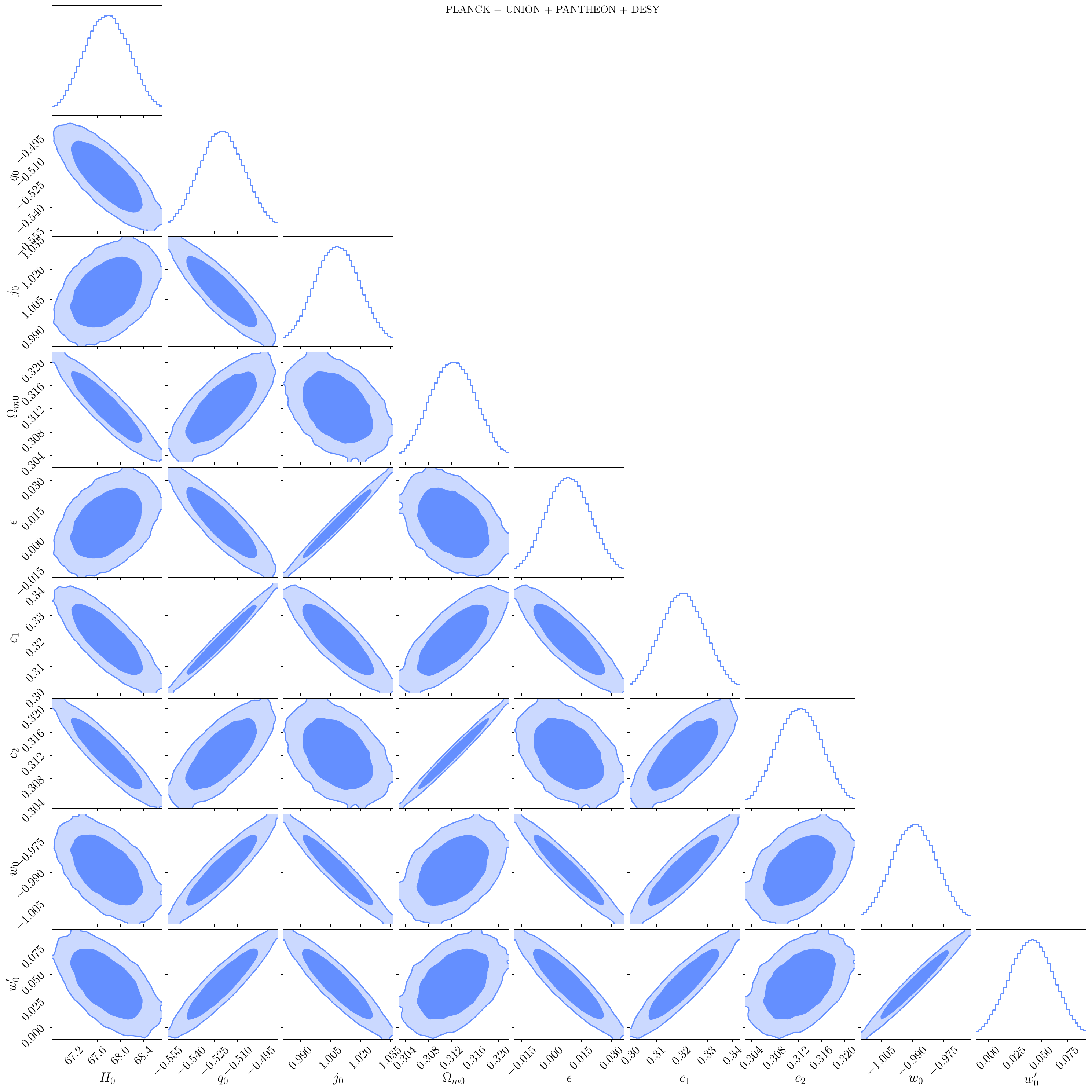}
        \caption{Model II}
    \end{subfigure}
    \hfill
    \begin{subfigure}[b]{0.4\linewidth}
        \includegraphics[width=\linewidth]{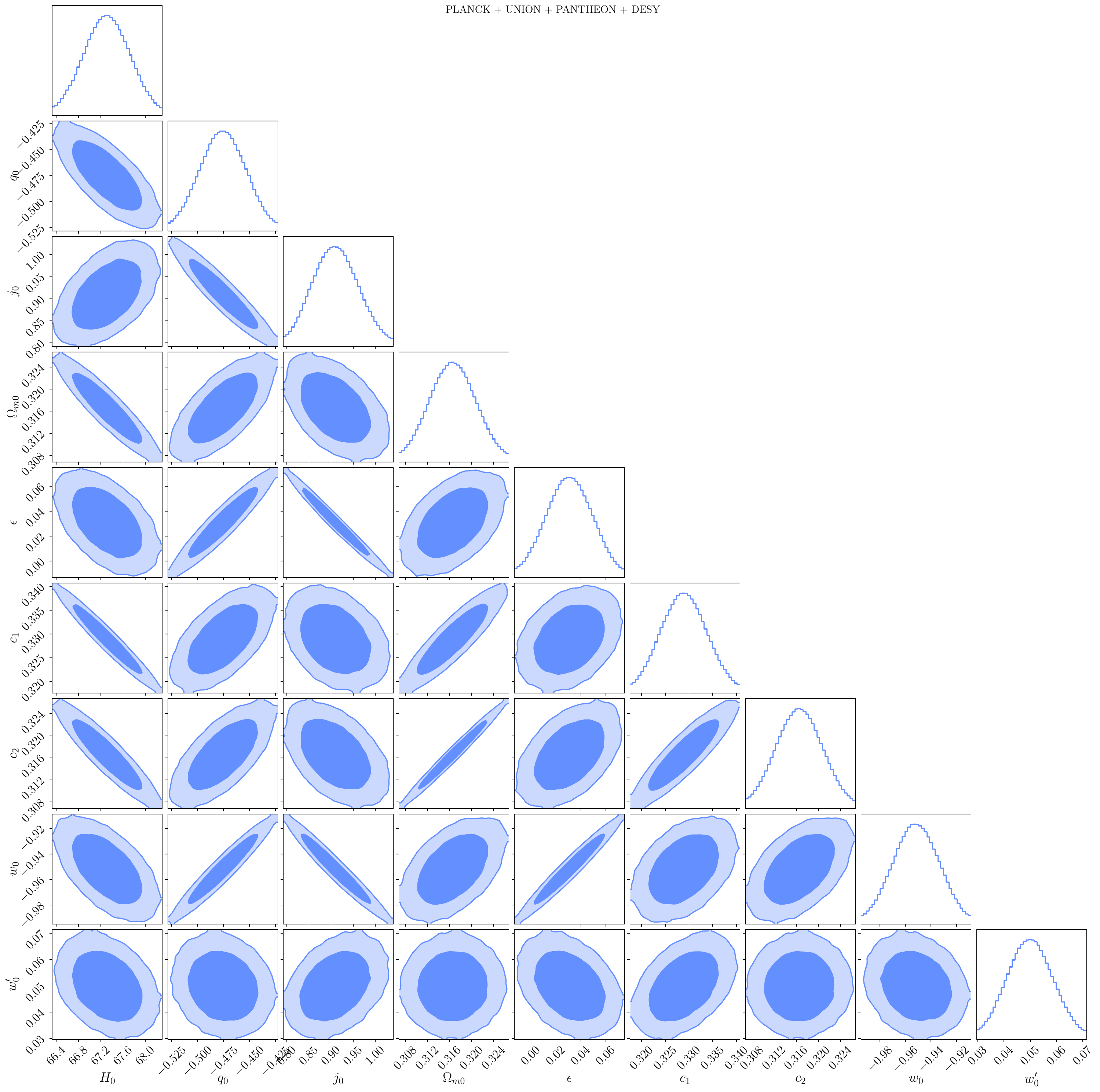}
        \caption{Model III}
    \end{subfigure}
    \caption{Confidence contours for the parameters $\{H_0, q_0, j_0, \Omega_{m0}, \epsilon, c_1, c_2, w_{\mathrm{DE}}(z)\vert_{z=0}, w'_{\mathrm{DE}}(z)\vert_{z=0} \}$ at 68\% and 95\% confidence levels for each model obtained with the DESI + Union3 + Pantheon+ + DESY5 + Planck dataset combination.}
    \label{fig:corner}
\end{figure}
\pagebreak

\section{Evolution of various quantities for the three evolutionary models considered}\label{app:B}

\begin{enumerate}

\item {\bf Model I: $j=1+3\epsilon(q+1)$}

\begin{subequations}
    \begin{eqnarray}
        h^2(z) &=& \frac{c_1 + 2(1+z)^{3 \epsilon +3}}{c_1+2}\,,
        \\
        q(z) &=& \frac{(3 \epsilon +1) (z+1)^{3 \epsilon +3}-c_1}{c_1+2 (z+1)^{3 \epsilon +3}}\,,
        \\
        j(z) &=& 3 \epsilon  \left[\frac{(3 \epsilon +1) (z+1)^{3 \epsilon +3}-c_1}{c_1+2 (z+1)^{3 \epsilon +3}}+1\right]+1\,,
        \\
        \Omega_m(z) &=& \frac{c_2 (z+1)^3}{c_1+2 (z+1)^{3 \epsilon +3}}\,,
        \\
        w_{\rm DE}(z) &=& \frac{2 \epsilon  (z+1)^{3 \epsilon +3}-c_1}{c_1+(z+1)^3 \left(2 (z+1)^{3 \epsilon }-c_2\right)} \,.
    \end{eqnarray}
\end{subequations}
Solving the equations
\begin{equation}
    \{q(0),j(0),\Omega_{m}(0)\}=\{q_0,j_0,\Omega_{m0}\}
\end{equation}
simultaneously we get
\begin{equation}\label{const_redef-I}
\epsilon = \frac{j_0-1}{3 (q_0+1)},\quad c_1 = \frac{j_0-2q_0^2-q_0}{(q_0+1)^2},\quad c_2 = \frac{\Omega_{m0}(j_0+3q_0+2)}{(q_0+1)^2},
\end{equation}
so that 
\begin{subequations}\label{cosmology-I}
    \begin{eqnarray}
        h^2(z) &=& \frac{2 (1 + z)^{(2 + j_0 + 3 q_0)/(1 + q_0)} + [j_0 - q_0 (1 + 2 q_0)]/(1 + q_0)^2}{(2 + j_0 + 3 q_0)/(1 + q_0)^2}\,,
        \\
        q(z) &=& \frac{(q_0+1) (j_0+q_0) (z+1)^{\frac{j_0+3 q_0+2}{q_0+1}}-j_0+2 q_0^2+q_0}{(q_0+1)^2 \left[2 (z+1)^{\frac{j_0+3 q_0+2}{q_0+1}}+\frac{j_0-q_0 (2 q_0+1)}{(q_0+1)^2}\right]}\,,
        \\
        j(z) &=& \frac{(j_0-1) (j_0+3 q_0+2) (z+1)^{\frac{j_0+3 q_0+2}{q_0+1}}}{2 (q_0+1)^2 (z+1)^{\frac{j_0+3 q_0+2}{q_0+1}}+j_0-q_0 (2 q_0+1)}+1\,,
        \\
        \Omega_m(z) &=& \frac{\Omega_{m0} (z+1)^3 (j_0+3 q_0+2)}{(q_0+1)^2 \left[2 (z+1)^{\frac{j_0+3 q_0+2}{q_0+1}}+\frac{j_0-q_0 (2 q_0+1)}{(q_0+1)^2}\right]}\,,
        \\
        w_{\rm DE}(z) &=& \frac{(q_0+1)^2 \left[\frac{-j_0+2 q_0^2+q_0}{(q_0+1)^2}+\frac{2 (j_0-1) (z+1)^{\frac{j_0+3 q_0+2}{q_0+1}}}{3 (q_0+1)}\right]}{(z+1)^3 \left[2 (q_0+1)^2 (z+1)^{\frac{j_0-1}{q_0+1}}-\Omega_{m0} (j_0+3 q_0+2)\right]+j_0-2 q_0^2-q_0}\,.\label{wDE_cosmology-I}
    \end{eqnarray}
\end{subequations}

\item {\bf Model II: $j=1+\epsilon$}

\begin{subequations}
    \begin{eqnarray}
    h^2(z) &=& c_1(1+z)^{\frac{3+\sqrt{9+8\epsilon}}{2}}+(1-c_1)(1+z)^{\frac{3-\sqrt{9+8\epsilon}}{2}}\,,
        \\
    q(z) &=& \frac{-1-\sqrt{9+8\epsilon}+c_1\left(1-(1+z)^{\sqrt{9+8\epsilon}}+\sqrt{9+8\epsilon}+(1+z)^{\sqrt{9+8\epsilon}}\sqrt{9+8\epsilon}\right)}{4+4c_1\left(-1+(1+z)^{\sqrt{9+8\epsilon}}\right)}\,,\nonumber
        \\
    &&    \\
    j(z) &=& 1+\epsilon\,,
        \\
   \Omega_m(z) &=& \frac{c_2(1+z)^{\frac{\sqrt{9+8\epsilon}+3}{2}}}{1+c_1\left[-1+(1+z)^{\sqrt{9+8\epsilon}}\right]}\,,
        \\
    w_{\rm DE}(z) &=& \frac{c_1 \left[\sqrt{8 \epsilon +9} (z+1)^{\sqrt{8 \epsilon +9}}-3 (z+1)^{\sqrt{8 \epsilon +9}}+\sqrt{8 \epsilon +9}+3\right]-\sqrt{8 \epsilon +9}-3}{6 c_1 \left[(z+1)^{\sqrt{8 \epsilon +9}}-1\right]-6 c_2 (z+1)^{\frac{1}{2} \left(\sqrt{8 \epsilon +9}+3\right)}+6}\,.
    \end{eqnarray}
\end{subequations}
Solving the equations
\begin{equation}
    \{q(0),j(0),\Omega_{m}(0)\}=\{q_0,j_0,\Omega_{m0}\}
\end{equation}
simultaneously we get
\begin{equation}\label{const_redef-II}
\epsilon = -1+j_0\,,\quad c_1 = \frac{4 \sqrt{8 j_0+1} q_0+8 j_0+\sqrt{8 j_0+1}+1}{2 (8 j_0+1)},\quad c_2 = \Omega_{m0}\,,
\end{equation}
so that
\begin{subequations}\label{cosmology-II}
    \begin{eqnarray}
        h^2(z) &=& (z+1)^{\frac{3}{2}-\frac{1}{2} \sqrt{8 j_0+1}} \left\{\frac{\left(\sqrt{8 j_0+1}+4 q_0+1\right) \left[(z+1)^{\sqrt{8 j_0+1}}-1\right]}{2 \sqrt{8 j_0+1}}+1\right\}\,,
        \\
        q(z) &=& \frac{\left(\sqrt{8 j_0+1}-1\right) q_0 (z+1)^{\sqrt{8 j_0+1}}+\left(\sqrt{8 j_0+1}+1\right) q_0+2 j_0 \left[(z+1)^{\sqrt{8 j_0+1}}-1\right]}{(4 q_0+1) (z+1)^{\sqrt{8 j_0+1}}+\sqrt{8 j_0+1} (z+1)^{\sqrt{8 j_0+1}}+\sqrt{8 j_0+1}-4 q_0-1}\,,\nonumber
        \\
        && \\
        j(z) &=& j_0\,,
        \\
        \Omega_m(z) &=& \frac{\Omega_{m0} (z+1)^{\frac{1}{2} \left(\sqrt{8 j_0+1}+3\right)}}{\frac{\left(\sqrt{8 j_0+1}+4 q_0+1\right) \left((z+1)^{\sqrt{8 j_0+1}}-1\right)}{2 \sqrt{8 j_0+1}}+1}\,,
        \\
        w_{\rm DE}(z) &=& \frac{\frac{\left(4 \sqrt{8 j_0+1} q_0+8 j_0+\sqrt{8 j_0+1}+1\right) \left(\sqrt{8 j_0+1} (z+1)^{\sqrt{8 j_0+1}}-3 (z+1)^{\sqrt{8 j_0+1}}+\sqrt{8 j_0+1}+3\right)}{16 j_0+2}-\sqrt{8 j_0+1}-3}{\frac{3 \left(\sqrt{8 j_0+1}+4 q_0+1\right) \left((z+1)^{\sqrt{8 j_0+1}}-1\right)}{\sqrt{8 j_0+1}}-6 \Omega_{m0} (z+1)^{\frac{1}{2} \left(\sqrt{8 j_0+1}+3\right)}+6}\,.\nonumber
        \\
        && \label{wDE_cosmology-II}
    \end{eqnarray}
\end{subequations}

\item {\bf Model III: $j=1+3\epsilon\left(q-\frac{1}{2}\right)$}

\begin{subequations}
    \begin{eqnarray}
    h^2(z) &=& c_1(1+z)^3+(1-c_1)(1+z)^{3\epsilon}\,,
        \\
    q(z) &=& \frac{1}{2}\frac{c_1(1+z)^3-(1-c_1)(2-3\epsilon)(1+z)^{3\epsilon}}{c_1(1+z)^3+(1-c_1)(1+z)^{3\epsilon}}\,,\nonumber
        \\
    &&    \\
    j(z) &=& \frac{1}{2}\frac{c_1(1+z)^3+(1-c_1)[2-9\epsilon(1-\epsilon)](1+z)^{3\epsilon}}{c_1(1+z)^3+(1-c_1)(1+z)^{3\epsilon}}\,,
        \\
   \Omega_m(z) &=& \frac{c_2(1+z)^3}{c_1(1+z)^3+(1-c_1)(1+z)^{3\epsilon}}\,,
        \\
    w_{\rm DE}(z) &=& -\frac{(1-c_1)(1-\epsilon)}{(1-c_1)+(c_1-c_2)(1+z)^{3-3\epsilon}}\,.
    \end{eqnarray}
\end{subequations}
Solving the equations
\begin{equation}
    \{q(0),j(0),\Omega_{m}(0)\}=\{q_0,j_0,\Omega_{m0}\}
\end{equation}
simultaneously we get
\begin{equation}\label{const_redef-III}
\epsilon = \frac{2(1-j_0)}{3(1-2q_0)}\,,\quad c_1 = \frac{2(j_0-q_0-2q_0^2)}{1-6q_0+2j_0},\quad c_2 = \Omega_{m0}\,,
\end{equation}
so that
\begin{subequations}\label{cosmology-III}
    \begin{eqnarray}
        h^2(z) &=& \frac{(1-2 q_0)^2 (z+1)^{\frac{2 (j_0-1)}{2 q_0-1}}+2 (z+1)^3 (j_0-q_0 (2 q_0+1))}{2 j_0-6 q_0+1}\,,
        \\
        q(z) &=& \frac{(2 q_0-1) (j_0-2 q_0) (z+1)^{\frac{2 (j_0-1)}{2 q_0-1}}+(z+1)^3 (j_0-q_0 (2 q_0+1))}{(1-2 q_0)^2 (z+1)^{\frac{2 (j_0-1)}{2 q_0-1}}+2 (z+1)^3 (j_0-q_0 (2 q_0+1))}\,,
        \\
        j(z) &=& -\frac{-\left((j_0-2 q_0) (2 j_0-2 q_0-1) (z+1)^{\frac{2 (j_0-1)}{2 q_0-1}}\right)-2 (z+1)^3 (j_0-q_0 (2 q_0+1))}{(1-2 q_0)^2 (z+1)^{\frac{2 (j_0-1)}{2 q_0-1}}+2 (z+1)^3 (j_0-q_0 (2 q_0+1))}\,,
        \\
        \Omega_m(z) &=& -\frac{\Omega_{m0} (2 j_0-6 q_0+1)}{-(1-2 q_0)^2 (z+1)^{\frac{2 (j_0-1)}{2 q_0-1}-3}-2 j_0+2 q_0 (2 q_0+1)}\,,
        \\
        w_{\rm DE}(z) &=& \frac{2 j_0-6 q_0+1}{(6 q_0-3) \left(1-\frac{(2 j_0 (\Omega_{m0}-1)+2 q_0 (2 q_0-3 \Omega_{m0}+1)+\Omega_{m0}) (z+1)^{3-\frac{2 (j_0-1)}{2 q_0-1}}}{(1-2 q_0)^2}\right)}\,.\label{wDE_cosmology-III}
    \end{eqnarray}
\end{subequations}
\end{enumerate}

It can be noted that for all three evolutionary models considered, we have
\begin{equation}
    w_{\text{DE}}(0) = \frac{1 - 2q_0}{3(\Omega_{m0} - 1)}\,.
\end{equation}

\end{widetext}

\clearpage

\bibliography{references}

\end{document}

%% file: preamble.tex
\usepackage[english]{babel}

\usepackage{mathrsfs}
\usepackage{amsfonts}
\usepackage{amsmath}
\usepackage{amsthm}
\usepackage{amssymb}
\usepackage{physics}
\usepackage{dsfont}
\usepackage{esint}

\usepackage{enumerate}
\usepackage[shortlabels]{enumitem}
\usepackage{framed}
\usepackage{csquotes}

\usepackage{tabularx}
\usepackage{xcolor}
\usepackage{minted}

\usepackage{subcaption}
\usepackage{tikz}
\usetikzlibrary{automata, positioning}

\DeclareUnicodeCharacter{223C}{$\sim$}



%% file: format.tex
\usepackage{titlesec}
\usepackage[many]{tcolorbox}

\definecolor{LightGray}{gray}{0.95}
\definecolor{DarkGray}{gray}{0.3}

\newtcbtheorem[auto counter, number within=section]{theorem}{Theorem}
    {enhanced,
    colback = LightGray,
    colbacktitle = LightGray,
    coltitle = black,
    boxrule = 0pt,
    frame hidden,
    borderline west = {0.0mm}{0.0mm}{black},
    fonttitle = \bfseries\sffamily,
    breakable,
    before skip = 2ex,
    after skip = 2ex
}{theorem}

\newtcbtheorem[auto counter, number within=section]{example}{Example}
    {enhanced,
    colback = LightGray,
    colbacktitle = DarkGray,
    coltitle = white,
    boxrule = 1.5pt,
    colframe = DarkGray,
    fonttitle = \bfseries\sffamily,
    before skip = 2ex,
    after skip = 2ex
}{example}

\tcbuselibrary{skins, breakable, minted}

\newtcblisting[auto counter, number within=section]{pythoncode}[1]{
    listing only,
    listing engine=minted,
    minted language=python,
    title={Python code \thetcbcounter: #1}, 
    coltitle=LightGray,
    fonttitle=\bfseries\sffamily,
    boxrule = 1.5pt,
    before skip=2ex,
    after skip=2ex,
    enhanced,
    top=0pt,
    bottom=0pt,
    boxsep=1pt,
    minted options={
        baselinestretch=1,
        bgcolor=LightGray,
        fontsize=\small,
        linenos,
        numbersep=20pt, 
        xleftmargin=-5pt 
    }
}

%% file: commands.tex
\makeatletter
\newcommand*\bigcdot{\mathpalette\bigcdot@{.5}}
\newcommand*\bigcdot@[2]{\mathbin{\vcenter{\hbox{\scalebox{#2}{$\m@th#1\bullet$}}}}}
\makeatother

